\documentclass[aps,prl,reprint,amsmath,amssymb,superscriptaddress,showpacs,floatfix]{revtex4-2}
\usepackage{graphicx}
\usepackage{dcolumn}
\usepackage{bm}
\usepackage[colorlinks=true,citecolor=blue,linkcolor=magenta, breaklinks=true]{hyperref}
\usepackage[usenames, dvipsnames]{color}
\usepackage{float}

\usepackage{lipsum}
\usepackage{titlesec}
\titlespacing\section{0pt}{12pt plus 4pt minus 4pt}{1pt plus 20pt minus 2pt}
\usepackage{xcolor}
\usepackage{tabularx}
\usepackage{amsmath}
\usepackage{comment}
\usepackage{afterpage}
\usepackage{placeins}
\usepackage{booktabs}
\usepackage{array}
\usepackage{setspace}
\graphicspath{{Figs/}}
\usepackage{hhline}
\usepackage{xfrac}
\usepackage{float,graphicx}
\usepackage{mathtools}
\usepackage{listings}
\usepackage{amssymb}
\usepackage{titlesec}
\usepackage{amsfonts}

\usepackage{epstopdf}
\catcode`@11
\def\seceqaa{\@addtoreset{equation}{section}
\def\theequation{A\arabic{equation}}}
\def\seceqbb{\@addtoreset{equation}{section}
\def\theequation{B\arabic{equation}}}
\def\seceqcc{\@addtoreset{equation}{section}
\def\theequation{C\arabic{equation}}}
\def\seceqdd{\@addtoreset{equation}{section}
\def\theequation{D\arabic{equation}}}
\def\seceqee{\@addtoreset{equation}{section}
\def\theequation{E\arabic{equation}}}
\def\seceqff{\@addtoreset{equation}{section}
\def\theequation{F\arabic{equation}}}
\def\seceqgg{\@addtoreset{equation}{section}
\def\theequation{G\arabic{equation}}}
\def\seceqhh{\@addtoreset{equation}{section}
\def\theequation{H\arabic{equation}}}
\catcode`@11

\begin{document}

\title{
	Polarization-Driven Charge Frustration and Emergent Phases\\ in the One-Dimensional Extended Hubbard Model
} 

\author{Sourabh Saha}
\affiliation{Department of Condensed Matter and Materials Physics,
S. N. Bose National Centre for Basic Sciences, JD Block, Sector III, Salt Lake, Kolkata 700106, India}
\author{Jeroen van den Brink}
\affiliation{Institute for Theoretical Solid State Physics, IFW Dresden, 01069 Dresden, Germany}
\affiliation{Würzburg-Dresden Cluster of Excellence ct.qmat, 97074 Würzburg, Germany}
\affiliation{Department of Physics, Technical University Dresden, 01069 Dresden, Germany}
\author{Manoranjan Kumar}
\email{manoranjan.kumar@bose.res.in}
\affiliation{Department of Condensed Matter and Materials Physics,
S. N. Bose National Centre for Basic Sciences, JD Block, Sector III, Salt Lake, Kolkata 700106, India}
\author{Satoshi Nishimoto}
\email{s.nishimoto@ifw-dresden.de}
\thanks{Last two authors contributed equally to this work.}
\affiliation{Institute for Theoretical Solid State Physics, IFW Dresden, 01069 Dresden, Germany}
\affiliation{Department of Physics, Technical University Dresden, 01069 Dresden, Germany}

\date{\today}

\begin{abstract}
Frustration is a key driver of exotic quantum phases, yet its role in charge dynamics remains largely unexplored. We show that charge frustration — induced by electronic polarization effects — stabilizes unconventional insulating states in the one-dimensional extended Hubbard model. Using exact diagonalization and density-matrix renormalization group, we uncover a charge-disordered phase that remains insulating despite lacking long-range order and possessing an effectively attractive on-site interaction — a behavior reminiscent of gapful spin liquids in frustrated spin systems. We also identify a fragile ferroelectric phase and a charge-density-wave state with emergent eight-site periodicity. These findings establish charge frustration, driven by charge-dipole interactions, as a robust mechanism for realizing exotic phases in low-dimensional correlated systems, with implications for organic conductors, transition-metal oxides, and ultracold polar molecules.
\end{abstract}

\maketitle

{\it Introduction.}---
Strongly correlated electron systems exhibit a wide range of quantum phenomena, including superconductivity, magnetism, and charge ordering~\cite{dagotto2005complexity,lee2006doping}. The Hubbard model and its extended variants have long served as fundamental theoretical frameworks for capturing these phenomena in correlated
systems~\cite{qin2022hubbard,arovas2022hubbard}. While the effects of Coulomb interactions--both short-range and long-range--are well studied, the role of polarization effects, particularly charge-dipole interactions, remains relatively underexplored despite their relevance in real materials and experimental systems.

Charge-dipole interactions, which couple mobile charges to induced dipoles, introduce qualitatively new effects distinct from purely Coulombic interactions. These interactions become especially important in low-dimensional systems such as organic conductors, transition-metal oxides, and ultracold polar molecules~\cite{mosadeq2015quantum,talebi2019spin,huang2009quantum}. Recent experiments with dipolar Fermi gases, including ${}^{161}$Dy~\cite{lu2012quantum} and polar molecules like ${}^{23}$Na${}^{40}$K and ${}^{40}$K${}^{87}$Rb~\cite{cheng2012ultracold,chotia2012long,ni2008high}, further underscore the need for a systematic theoretical understanding of polarization-driven phenomena.

One-dimensional (1D) systems strongly enhance quantum fluctuations, making them highly sensitive to competing interactions. Such competition frequently gives rise to exotic ground states, including charge-density waves (CDW), bond-order waves (BOW), spin-density waves (SDW), superconductivity, and phase separation~\cite{nakamura1999mechanism,nakamura2000tricritical,sengupta2002bond,eric2002Ground,tsuchiizu2002Phase,dhar2016population,kumar2009tuning,kumar2010bond,ejima2007phase,qu2022spin}. Despite extensive studies of spin frustration, charge frustration explicitly induced by charge-dipole couplings has remained largely unexplored, apart from some pioneering works~\cite{van1995new,meinders1995atomic,torio2000phase}. This gap poses fundamental questions about the nature and stability of emergent phases in low-dimensional correlated electron systems.

In this Letter, we present a systematic study of the ground-state properties of the 1D extended Hubbard model explicitly incorporating charge-dipole interactions. Using exact diagonalization (ED) and density-matrix renormalization group (DMRG) methods~\cite{white1992dmrg}, we reveal rich phase diagrams featuring unconventional insulating phases. Notably, we demonstrate that charge frustration driven by the competition between Coulomb and charge-dipole interactions stabilizes novel states: a fragile ferroelectric (FE) phase, a previously unreported charge-disordered insulating (CDI) phase, and a remarkable CDW phase with an emergent eight-site periodicity. Our results highlight charge frustration, driven by polarization effects, as a new route to exotic insulating and charge-ordered phases in low-dimensional correlated systems, with relevance to organic conductors, transition-metal oxides, and ultracold polar molecules.

{\it Model.}---
We consider the 1D extended Hubbard model incorporating charge-dipole interactions to account for polarization effects. The Hamiltonian is
\begin{eqnarray}
	H = -t \sum_{i,\sigma} \left( c_{i,\sigma}^\dagger c_{i+1,\sigma} + \text{H.c.} \right) + (U - 4P) \sum_{i} n_{i\uparrow} n_{i\downarrow} \nonumber \\
	+ V \sum_{i} n_i n_{i+1} + 2P \sum_{i} n_i n_{i+2} - 2P \sum_{i} n_i, \hspace{0.5cm}
	\label{eq1}
\end{eqnarray}  
where $c_{i,\sigma}^\dagger$ ($c_{i,\sigma}$) creates (annihilates) an electron with spin $\sigma$ at site $i$, and $n_i=n_{i,\uparrow}+n_{i,\downarrow}$. We focus on the half-filled case. The parameters $t$, $U$, and $V$ represent the nearest-neighbor hopping amplitude, on-site Coulomb repulsion, and nearest-neighbor Coulomb interaction, respectively. The polarization energy $P$ modifies the interactions in two distinct ways: it reduces the on-site repulsion by $4P$ and introduces a next-nearest-neighbor repulsion of strength $2P$. A detailed derivation of the charge-dipole interaction is provided in the Supplemental Material \cite{suppmat}.

\begin{figure}[t!]
    \centering
\includegraphics[width=0.7\linewidth]{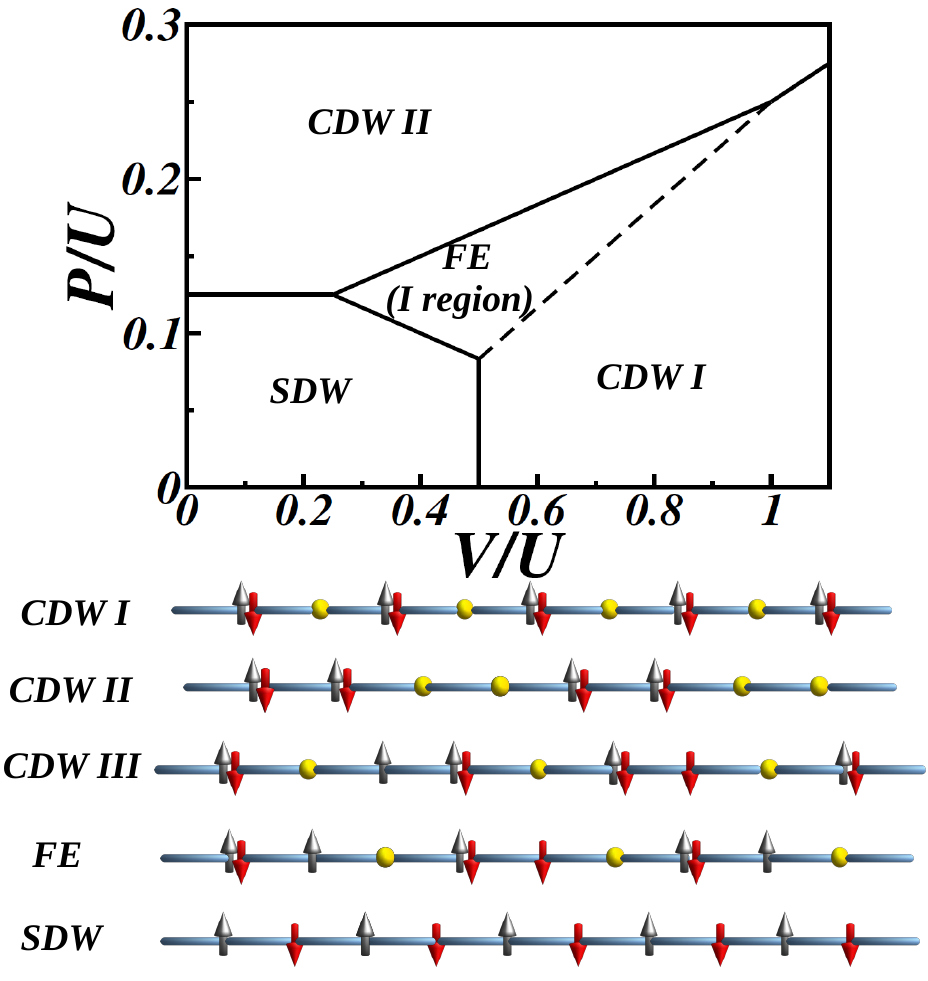}
    \caption{Phase diagram of the system \eqref{eq1} in the atomic limit ($t=0$) within the ($V/U$, $P/U$) parameter space. The schematic electron configurations for each phase are illustrated, assuming an infinitesimally small $t$ to indicate the corresponding spin configurations. The FE, CDW I, and CDW III states are energetically degenerate along the dashed line.}
    \label{fig1}
\end{figure}


{\it Phases in the atomic limit.}---
To gain initial insight into the ground-state properties of the model, we consider the atomic limit ($t=0$). The phase diagram in this limit, shown in Fig.~\ref{fig1}, reveals one Mott insulating phase and three distinct CDW phases, each characterized by long-range charge order. The Mott insulator corresponds to a SDW state without long-range magnetic order; finite hopping introduces antiferromagnetic (AFM) fluctuations via an effective exchange interaction $\propto t^2/(U-P)$.

The three CDW phases exhibit twofold, threefold, and fourfold periodicities, denoted CDW I, FE, and CDW II, respectively, in line with previous studies~\cite{van1995new,meinders1995atomic,torio2000phase}. In addition, at the FE-CDW I phase boundary, an eightfold periodic CDW state - labeled CDW III - becomes degenerate (see the Supplemental Material \cite{suppmat}). This reflects a high degree of charge frustration near the phase boundaries, particularly in the region occupied by the FE phase.

This frustrated region is especially sensitive to quantum fluctuations when $t$ is finite. As we show below, hopping destabilizes the classical charge orders and gives rise to exotic quantum states. We refer to this fluctuation-prone sector as the {\it intermediate region} (I region) in the remainder of this paper.

{\it Methods.}---
We determine the phase boundaries using a level-crossing analysis~\cite{jullien1983finite,okamoto1992fluid} based on ED for several values of $U/t$. To refine these boundaries and explore regions inaccessible to the level-crossing approach, we perform large-scale DMRG simulations. Additional details are provided in the Supplemental Material \cite{suppmat}.

\begin{table}
	\caption{Sets of discrete symmetries used to classify excitation spectra at each phase transition in the level-crossing analysis.
	}
	\begin{center}
		\newcolumntype{P}{D{z}{-}{8}}
		\begin{tabular}{P|wc{2.5cm}wc{2.5cm}}
			\hline
			\hline
   \multicolumn{1}{c}{\rm transition} &
\multicolumn{2}{|c}{\rm symmetry quantum numbers} \\
			{\rm phase A}\,z\,{\rm phase B} & A:($\sigma$,$J$,$S$) & B:($\sigma$,$J$,$S$) \\
			\hline
			{\rm SDW}\,z\,{\rm CDW\,I}        &  (+1,+1,1)  &  (+1,-1,0)  \\
			{\rm SDW}\,z\,{\rm CDW\,II}       &  (+1,+1,1)  &  (+1,+1,0)  \\
			{\rm SDW}\,z\,{\rm I\,region}     &  (-1,+1,0)  &  (+1,-1,0)  \\
			{\rm SDW}\,z\,{\rm BOW}           &  (+1,+1,1)  &  (+1,+1,0)  \\
			{\rm CDW\,I}\,z\,{\rm BOW}        &  (+1,-1,0)  &  (-1,+1,0)  \\
			{\rm CDW\,I}\,z\,{\rm I\,region}  &  (+1,+1,0)  &  (+1,+1,1)  \\
			{\rm CDW\,II}\,z\,{\rm I\,region} &  (+1,+1,0)  &  (+1,+1,1)  \\
			\hline
			\hline
		\end{tabular}
	\end{center}
	\label{table:QNs}
\end{table}

\begin{figure}
    \centering
    \includegraphics[width=1.0\linewidth]{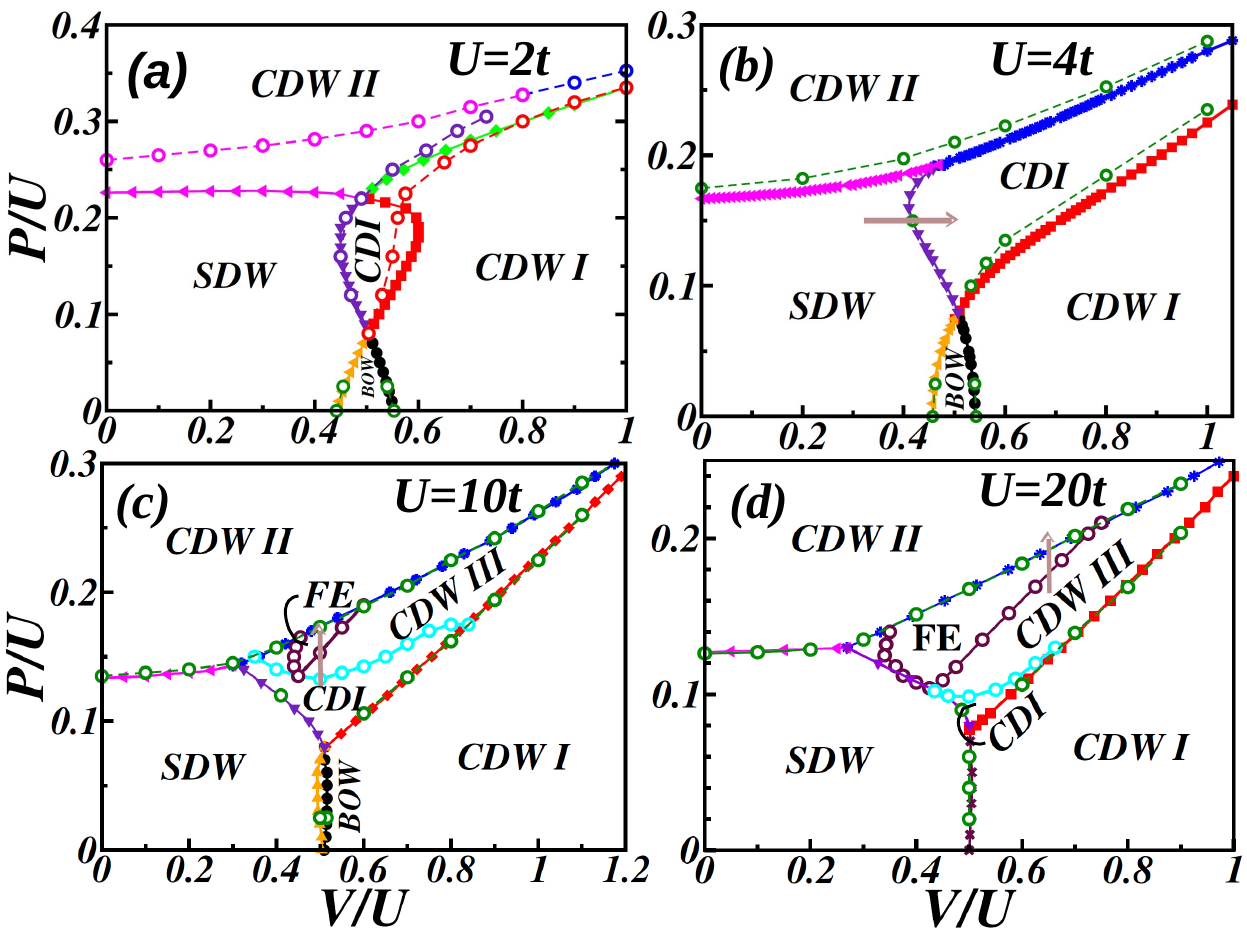}
    \caption{Ground-state phase diagrams of the model in Eq.~\eqref{eq1} for four values of $U/t$, plotted in the $(V/U, P/U)$ plane. Solid symbols denote phase boundaries determined via level-crossing analysis, while open symbols are obtained from DMRG. In (a), transitions of the same type are shown in the same color. Arrows in (b–d) indicate parameter paths used for analysis in subsequent sections.
    }
    \label{fig2}
\end{figure}

{\it Phase boundaries via level-crossing approach.}---
We estimate phase boundaries by analyzing level crossings in the excitation spectra. Each excitation is classified by discrete quantum numbers: inversion symmetry ($\sigma=\pm1$), electron-hole symmetry ($J=\pm1$), and total spin $S$. Transitions are identified by the crossing of energy levels with different symmetry labels, as summarized in Table~\ref{table:QNs}. In addition to the four insulating phases in Fig.~\ref{fig1}, this approach also captures the BOW phase, known to appear at weak to intermediate $U/t$~\cite{nakamura1999mechanism,nakamura2000tricritical}.

We thus distinguish five distinct phases: SDW, CDW I, CDW II, the I region, and BOW. The resulting phase diagrams for $U/t = 2, 4, 10$, and $20$, obtained from ED on a periodic $N = 12$ site cluster, are shown in Fig.~\ref{fig2}. The system size accommodates both the six-site periodicity of the FE state and the four-site periodicity of CDW II. Remarkably, the overall phase structure in the $(V/U, P/U)$ plane remains qualitatively similar down to $U=4t$. At $U=2t$, however, the I region narrows significantly. Notably, the BOW phase persists even in the presence of finite polarization. Phase boundaries and the stability of the BOW region are further refined using DMRG; see the Supplemental Material for details \cite{suppmat}.

Beyond the FE phase found in the atomic limit, we identify two additional insulating states within the I region: a CDI phase, which is neither charge-ordered nor Mott-type, and a long-period CDW phase (CDW III) exhibiting eight-site periodicity. These phases are challenging to detect via level-crossing due to symmetry constraints. We therefore perform a detailed analysis using DMRG, as described below.

\begin{figure}
    \centering
    \includegraphics[width=1.0\linewidth]{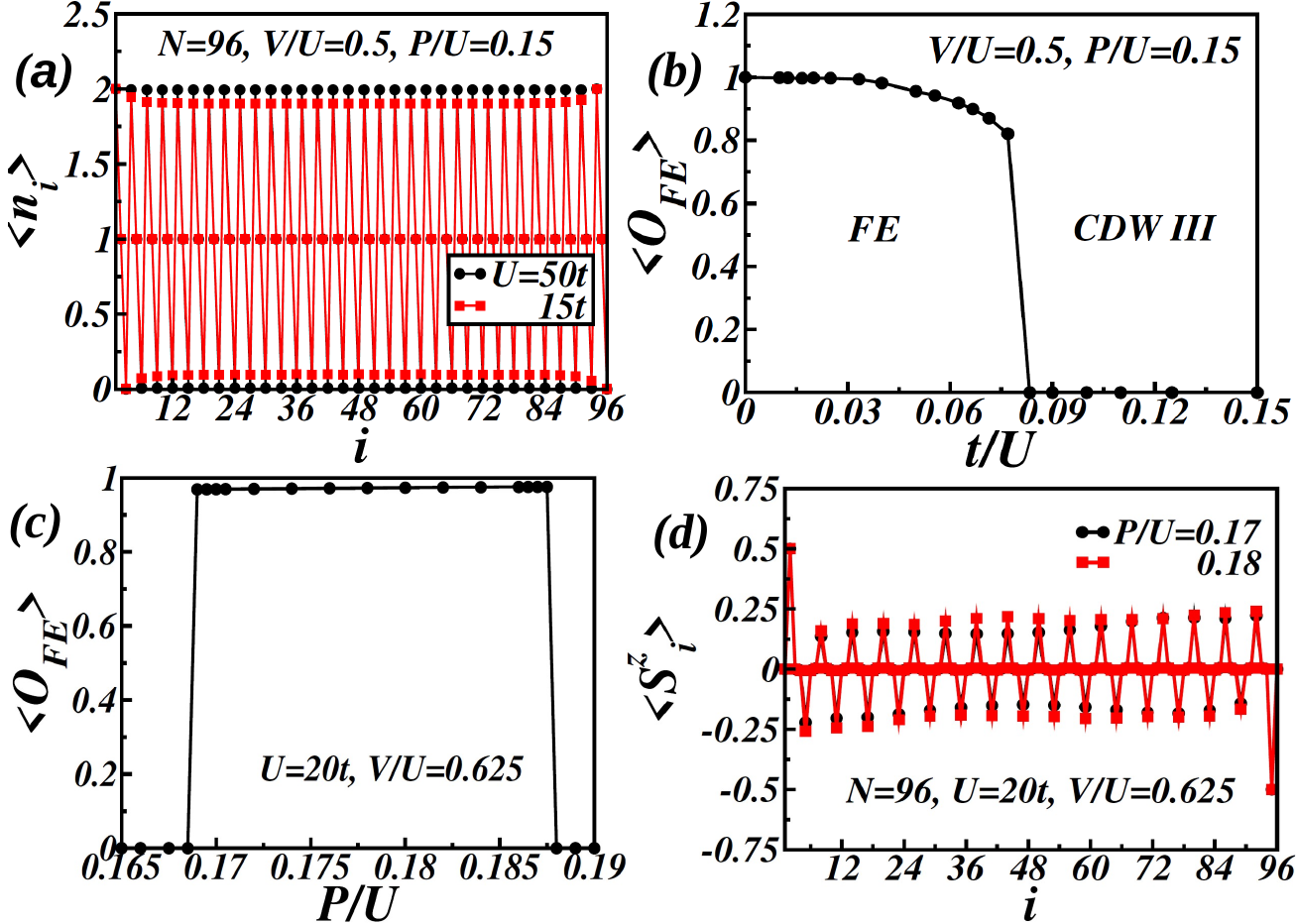}
    \caption{DMRG results for the FE state. (a) Charge density distribution at large and moderate $U/t$. (b, c) FE order parameter as a function of $t/U$ and $P/U$, respectively; the parameter path for (c) is shown in Fig.~\ref{fig2}(d). (d) Local spin profile $\langle S_i^z \rangle$ for representative parameters. All parameters are given in the figure.
    }
    \label{fig3}
\end{figure}

{\it Stability of FE state.}---
As noted, the entire I region corresponds to the FE phase in the atomic limit. However, this state is highly fragile against finite hopping and difficult to detect numerically~\cite{torio2000phase}. To probe its stability, we approach the FE state from large $U/t$. Figure~\ref{fig5}(a) shows the charge density for $U/t=50$ and $15$, at fixed $V/U=0.5$ and $P/U=0.15$, chosen near the FE phase center in the atomic limit. The characteristic three-site pattern - doubly occupied, singly occupied, and empty - is evident, corresponding to a CDW with wavevector $q=2\pi/3$. The modulation amplitude is reduced at $U/t=15$, reflecting enhanced quantum fluctuations.

To quantify long-range FE order, we compute the order parameter $\langle {\cal O}_{\rm FE} \rangle = \lim_{N \to \infty} \frac{1}{2}|\langle n_{N/2-2} \rangle - \langle n_{N/2} \rangle|$. Figure~\ref{fig5}(b) shows its dependence on $t/U$ at fixed $V/U=0.5$ and $P/U=0.15$. Starting from unity at $t/U = 0$, it decreases and vanishes abruptly only near $t/U \sim 0.08$, signaling a nearly first-order transition. This highlights the extreme fragility of the FE phase, which occupies a wide region at $U/t=20$ [Fig.\ref{fig2}(d)] but shrinks drastically at $U/t=10$ [Fig.\ref{fig2}(c)].

Figure~\ref{fig5}(c) further characterizes the transition by showing $\langle {\cal O}_{\rm FE} \rangle$ versus $P/U$ for $U/t=20$ and $V/U=0.625$. The system evolves from CDW I to FE to CDW III, with discontinuous jumps in $\langle {\cal O}_{\rm FE} \rangle$ at both boundaries, providing clear evidence for first-order transitions. This is consistent with the FE phase (threefold) mediating between two CDWs with coprime periodicities -- twofold (CDW I) and eightfold (CDW III).

Additional insight comes from the spin structure. Assuming superexchange between singly occupied sites, the FE phase maps to an AFM Heisenberg chain. Figure~\ref{fig5}(d) shows the $z$-component spin profile $\langle S^z_i \rangle$ for $P/U = 0.17$ and $0.18$ at $U/t = 20$, $V/U = 0.625$. As expected, $\langle S^z_i \rangle \approx 0$ at doubly occupied and empty sites, with staggered spin moments on singly occupied ones, schematically: ``$\cdots$\hspace{0.2cm}$\uparrow\downarrow$\hspace{0.2cm}$\uparrow$\hspace{0.2cm}$\circ$\hspace{0.2cm}$\uparrow\downarrow$\hspace{0.2cm}$\downarrow$\hspace{0.2cm}$\circ$\hspace{0.2cm}$\cdots$''. However, long-range magnetic order is absent due to the spin isotropy. This is confirmed by a vanishing spin gap. Note that the effective exchange interaction is extremely weak; for example, $J_{\rm eff} \sim 9.4 \times 10^{-4}$ for $U/t=20$, $V/U=0.625$, and $P/U=0.18$. Further details are provided in the Supplemental Material \cite{suppmat}.

Thus, we confirm that the I region hosts the FE phase at large $U/t$, but increasing hopping destabilizes it, giving rise to two novel phases (see below): a CDI without charge order and distinct from a Mott insulator, and a CDW III phase with long-range eight-site modulation ($q=3\pi/4$).

\begin{figure}[t]
    \centering
    \includegraphics[width=1.0\linewidth]{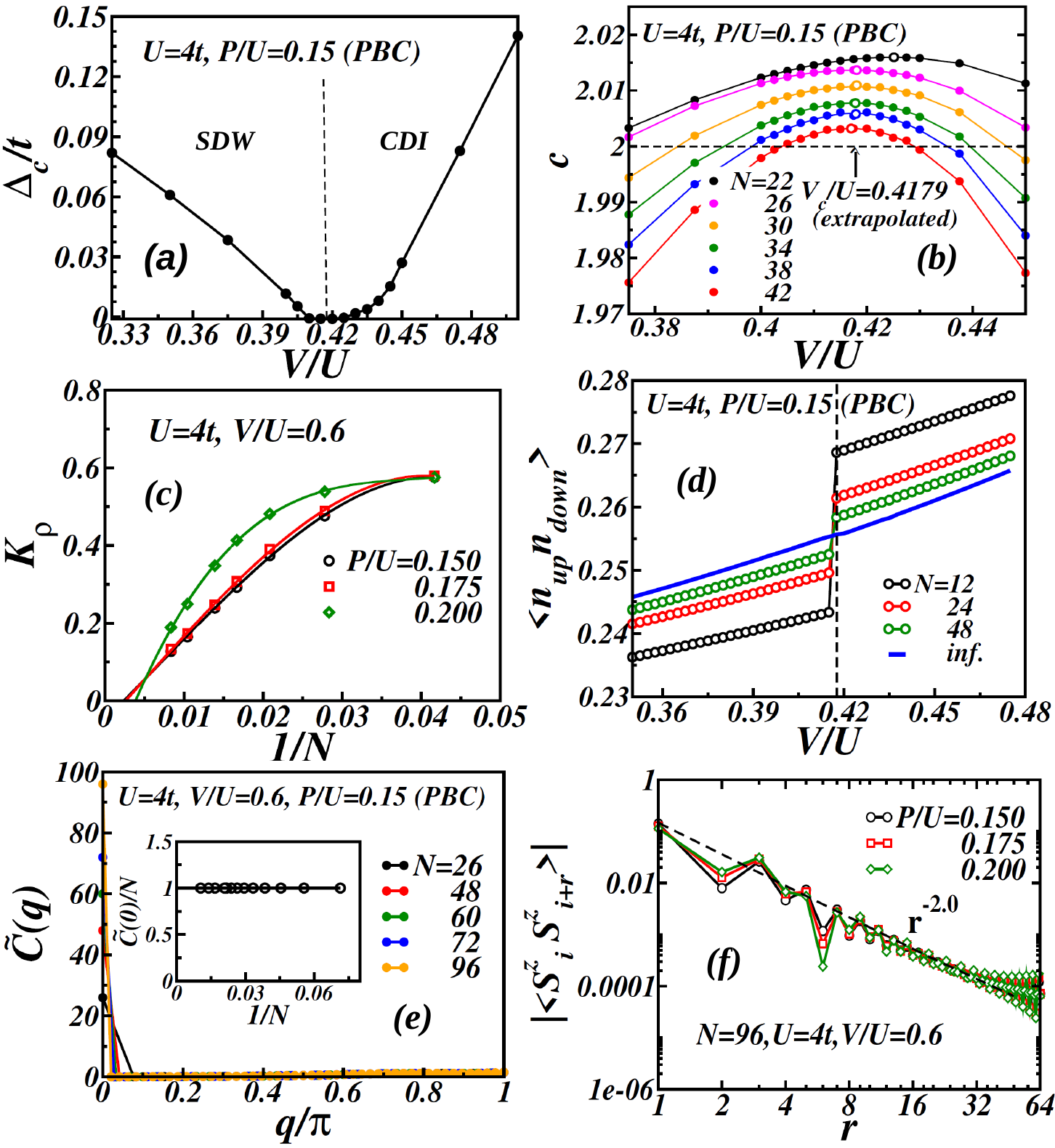}
    \caption{DMRG results for the SDW and CDI phases. (a) Charge gap $\Delta_{\rm c}/t$ as a function of $V/U$; the parameter path for (a,b,d) is shown in Fig.~\ref{fig2}(b). (b) Central charge $c$ versus $V/U$, with open circles marking the peak positions used for finite-size extrapolation. (c) Finite-size scaling analysis of the Tomonaga-Luttinger parameter $K_\rho$ within the CDI phase. (d) Double occupancy $\langle n_{i,\uparrow} n_{i,\downarrow} \rangle$ versus $V/U$, highlighting the transition from SDW to CDI phases. (e) Charge structure factor $\widetilde{C}(q)$ computed in the CDI phase; inset shows $\widetilde{C}(0)/N$ versus $1/N$. (f) Log-log plot of the spin-spin correlation function $|\langle S_i^z S_{i+r}^z \rangle|$ within the CDI phase. All parameters are given in the figure.
    }
    \label{fig4}
\end{figure}

{\it Properties of CDI phase.}---
We now examine the CDI phase in detail, focusing on its emergence from the SDW phase and its unconventional insulating character. Figure~\ref{fig4}(a) shows the charge gap, defined as $\Delta_c = \lim_{N\to\infty} \left[E_0(N+2,0)+E_0(N-2,0)-2E_0(N,0)\right]/2$, where $E_0(N_e,S^z)$ is the ground-state energy of a system with $N$ sites, $N_e$ electrons, and total spin-$z$ component $S^z$. The data are presented as a function of $V/U$ for $U/t=4$ and $P/U=0.15$. In the SDW phase, $\Delta_c$ is finite, as expected for a Mott insulator. As $V/U$ increases, the gap decreases, closes at the transition point, and reopens in the CDI phase, indicating a continuous transition between two charge-gapped states. Meanwhile, the spin gap remains zero throughout, as confirmed in the Supplemental Material \cite{suppmat}.

To further support these findings, we calculate the central charge, $c=3[S_{N}(N/2-1)-S_{N}(N/2)] / \ln[\cos(\pi/N)]$, where $S_N(l)$ is the von Neumann entanglement entropy of a subsystem of length $l$ in a periodic chain of length $N$~\cite{calabrese2004entanglement,PhysRevB.84.195108}. Figure~\ref{fig3}(b) shows $c$ as a function of $V/U$ for several system sizes at fixed $U/t=4$ and $P/U=0.15$. Each curve exhibits a single peak, and extrapolating these peaks to the thermodynamic limit locates the transition at $V/U=0.4179$, in excellent agreement with both the gap-closing point (0.4175) and the level-crossing estimate (0.4171). Notably, $c=2$ appears only at the critical point, where both spin and charge excitations are gapless, while $c=1$ holds in both the SDW and CDI phases due to a single gapless spin mode.

The insulating character of the CDI phase is further confirmed through the Tomonaga-Luttinger parameter $K_\rho$, defined via the slope of the charge structure factor $C(q)=\sum_{r} e^{iqr}(\langle n_i n_{i+r} \rangle-\langle n_i \rangle \langle n_{i+r} \rangle)$ in the limit of small momentum, $K_\rho = \pi \lim_{q \to 0} C(q)/q$~\cite{S.Ejima_2005}. As shown in Fig.~\ref{fig4}(c), $K_\rho$ extrapolates to zero throughout the CDI phase, confirming its incompressibility.

The CDI phase also exhibits unique local characteristics. Figure~\ref{fig3}(d) shows the double occupancy $\langle n_{i\uparrow} n_{i\downarrow} \rangle$, which remains below 0.25 in the SDW phase, consistent with repulsive interactions. However, it exceeds 0.25 in the CDI phase, indicating an effective on-site attraction. This distinguishes CDI from a conventional Mott insulator.

To examine the possibility of charge ordering or phase separation, we evaluate the structure factor without subtracting the mean density: $\widetilde{C}(q) = \sum_r e^{iqr} \langle n_i n_{i+r} \rangle$. Figure~\ref{fig3}(e) shows $\widetilde{C}(q)$ for $U/t=4$, $V/U =0.6$, and $P/U=0.15$. A single peak appears at $q=0$, with $\widetilde{C}(0)=N$, consistent with a uniform charge distribution. The absence of significant peaks at finite $q$ confirms that the system lacks long-range charge order, and the fact that $\widetilde{C}(0)$ does not exceed $N$ rules out phase separation. The charge profile thus remains spatially uniform, resembling that of a Mott insulator, though distinct in microscopic character.

Remarkably, despite the absence of long-range order, the CDI phase remains insulating. Its origin lies in the proximity to several competing charge-ordered phases - CDW I, CDW II, FE, and CDW III - which induce short-range patterns in the ground-state wavefunction. These local configurations act as effective double occupancy barriers, suppressing charge transport and localizing carriers. This mechanism is also reflected in the collapse of the charge dispersion in the single-particle excitation spectrum. Moreover, as shown in Fig.~\ref{fig3}(f) the spin–spin correlation function $|\langle S_i^z S_{i+r}^z \rangle|$ decays as $1/r^2$, indicating the absence of magnetic interactions. This rapid decay does not arise from conventional spin dynamics, but rather from the lack of effective spin couplings imposed by the underlying charge disorder. Additional supporting data are provided in the Supplemental Material \cite{suppmat}.

The CDI phase thus represents a novel type of insulating state - uniform, non-Mott, and stabilized by frustration - induced local correlations. A deeper theoretical understanding of its nature remains an open question for future investigation.

\begin{figure}[t]
    \centering
    \includegraphics[width=1.0\linewidth,height=9.5cm]{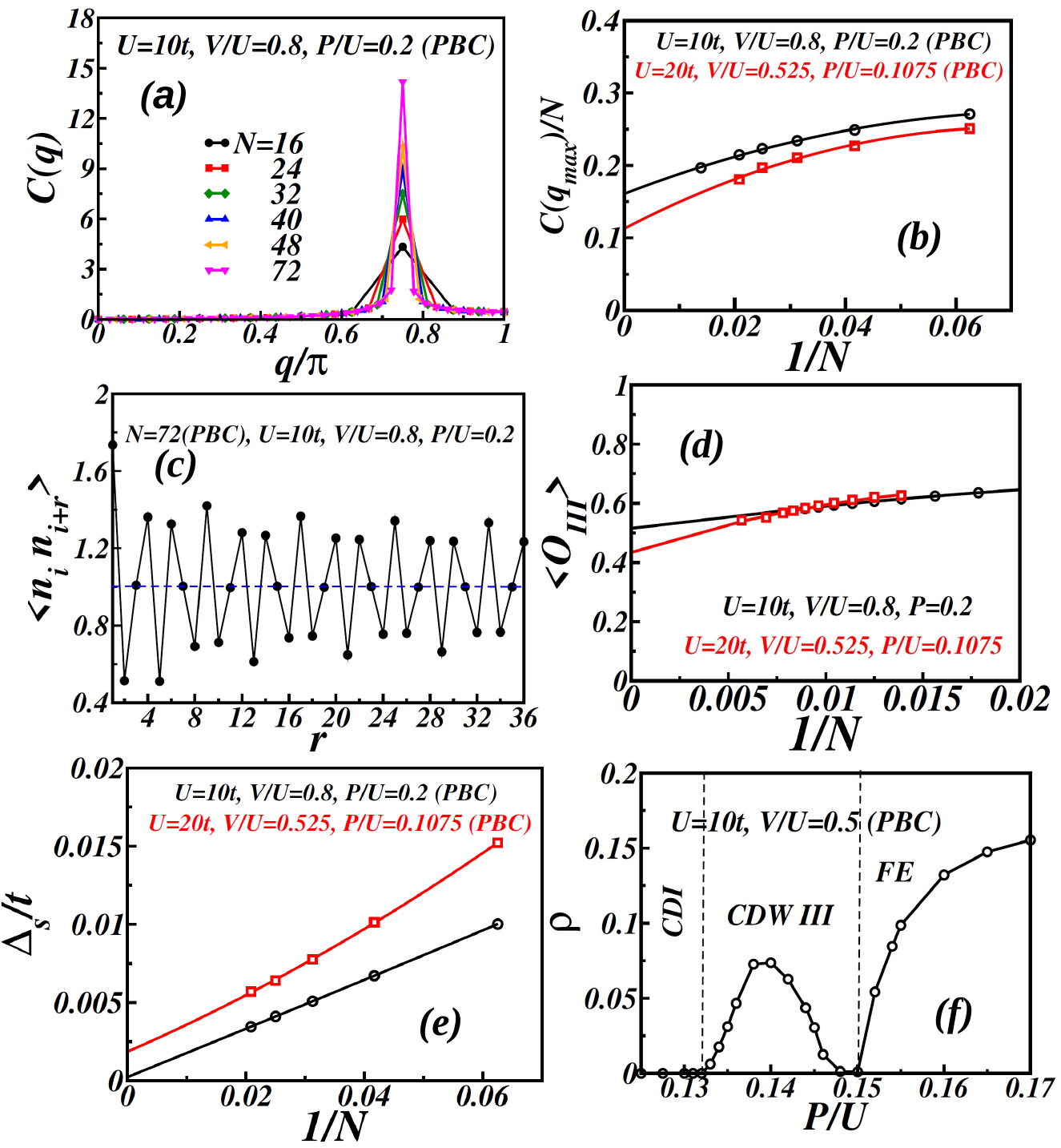}
    \caption{DMRG results for the CDW III phase. (a) Charge structure factor $C(q)$ and (b) finite-size scaling of the normalized peak height at $q=3\pi/4$. (c) Density-density correlations. (d) Finite-size scaling of the CDW III order parameter. (e) Finite-size scaling of the spin gap. (f) Charge order parameter $\rho$ versus $P/U$; the parameter path for (f) is shown in Fig.~\ref{fig2}(c). All parameters are given in the figure.
    }
    \label{fig5}
\end{figure}

{\it CDW state with eight-fold periodicity.}---
We now turn to the CDW III phase, which emerges within the I region at large $U/t$ [see Fig.~\ref{fig2}]. Figure~\ref{fig5}(a) shows the charge structure factor $C(q)$, computed at $U/t=10$, $V/U=0.8$, and $P/U=0.2$. A pronounced peak appears at $q=3\pi/4$, whose divergence with system size confirms long-range order with eight-site periodicity [Fig.~\ref{fig5}(b)].

To identify the charge alignment, we plot the density-density correlations in Fig.~\ref{fig5}(c), which reveal a characteristic pattern: ``$\cdots$\hspace{0.2cm}$\uparrow\downarrow$\hspace{0.2cm}$\circ$\hspace{0.2cm}$\uparrow$\hspace{0.2cm}$\uparrow\downarrow$\hspace{0.2cm}$\circ$\hspace{0.2cm}$\uparrow\downarrow$\hspace{0.2cm}$\downarrow$\hspace{0.2cm}$\circ$\hspace{0.2cm}$\cdots$''. Each period contains three charge-rich, three charge-poor, and two intermediate sites. To quantify this order, we define the CDW III order parameter as $\langle {\cal O}_{\rm III} \rangle = \frac{1}{6} \sum_{i=N/2-3}^{N/2+4} \gamma \langle n_i \rangle$, where $\gamma=1$ for charge-rich sites, $-1$ for charge-poor sites, and $0$ otherwise. Finite-size scaling confirms a nonzero value in the thermodynamic limit [Fig.~\ref{fig5}(d)], establishing the stability of the CDW III phase.

In the spin sector, this configuration localizes spin-1/2 moments on singly occupied sites, where charge patterns such as ``$\uparrow\downarrow$\hspace{0.2cm}$\circ$\hspace{0.2cm}$\uparrow\downarrow$'' and ``$\circ$\hspace{0.2cm}$\uparrow\downarrow$\hspace{0.2cm}$\circ$''appear alternately between them. This maps the system onto a spin-Peierls Heisenberg chain with alternating exchange couplings. In fact, from a small-cluster analysis, we estimate the effective couplings to be $J_1^{\rm eff}=0.01604t$ and $J_2^{\rm eff}=0.01422t$ for $U/t=20$, $V/U=0.525$, and $P/U=0.1075$, yielding a spin gap $\Delta_s/t \approx 0.0036$. This is in good agreement with the spin gap obtained directly from the original model, $\Delta_s = \lim_{N\to\infty}[E_0(N,1)-E_0(N,0)]$, which gives $\Delta_s \sim 0.0019t$ [Fig.~\ref{fig5}(e)]. Further details are provided in the Supplemental Material \cite{suppmat}.

To delineate the CDW III region, we compute the scaled structure factor $\rho = \lim_{N \to \infty} C(q_{\rm max})/N$ along a path from the CDI phase through CDW III into the FE phase at $U=10t$ and $V/U=0.5$ [Fig.~\ref{fig5}(f)]. In the CDI phase, $\rho=0$; it becomes finite in the CDW III phase at $q_{\rm max}=3\pi/4$, and vanishes again as the system transitions into the FE phase. There, a new peak emerges at $q_{\rm max}=2\pi/3$, signaling the onset of threefold FE charge order.

The CDW III phase originates near the CDW I–FE boundary at large $U/t$. In the atomic limit, its configuration is energetically degenerate with those of CDW I and FE on the boundary. Finite hopping lifts this degeneracy, and the gain in spin-Peierls energy can stabilize CDW III over FE. This mechanism explains why, in the $U/t=20$ phase diagram [Fig.~\ref{fig2}(d)], the region adjacent to CDW I - classically part of the FE phase - is replaced by CDW III.

{\it Summary.}---
We have investigated the ground-state properties of the 1D extended Hubbard model with charge-dipole interactions, using ED and DMRG calculations. The inclusion of polarization effects introduces competing interactions that drive strong charge frustration. Within this frustrated regime, we uncover a novel charge-disordered insulating phase that is charge-gapped yet non-Mott in character - lacking both charge order and magnetic interactions, and stabilized by local short-range correlations induced by nearby CDW states. We also identify the extreme fragility of the FE phase under quantum fluctuations, and a CDW phase with emergent eight-site periodicity. These phases arise from the interplay between charge frustration and hopping-induced quantum fluctuations. Our results shed light on fundamental mechanisms governing charge transport and coherence in low-dimensional systems, and offer insights relevant to the design of correlated materials and dipolar quantum platforms.


\section*{ACKNOWLEDGMENT}
We thank Ulrike Nitzsche for technical assistance and 
Masaaki Nakamura for fruitful discussion. S.S. acknowledges the financial support from Institute for Theoretical Solid State Physics of IFW Dresden and DST INSPIRE. This project was funded by the German Research Foundation (DFG) via the projects A05 of the Collaborative Research Center SFB 1143 (project-id 247310070) and through the W\"urzburg-Dresden Cluster of Excellence on Complexity and Topology in Quantum Matter ct.qmat (EXC 2147, Project No. 39085490).
\par

\medskip

\bibliography{reference}

\end{document}



\begin{center}
\large
{\bf Supplemental Material for\\
Polarization-Driven Charge Frustration and Emergent Phases\\ in the One-Dimensional Extended Hubbard Model}

\normalsize
\vspace{3.0mm}
Sourabh Saha$^1$, Jeroen van den Brink$^{2,3,4}$, Manoranjan Kumar$^{1,*}$, and Satoshi Nishimoto$^{2,4,\dagger}$

\small

\vspace{1.5mm}
{\it
$^1$Department of Condensed Matter and Materials Physics,
S. N. Bose National Centre for Basic Sciences,\\ JD Block, Sector III, Salt Lake, Kolkata 700106, India\\
$^2$Institute for Theoretical Solid State Physics, IFW Dresden, 01069 Dresden, Germany\\
$^3$W\"urzburg-Dresden Cluster of Excellence ct.qmat, 97074 W\"urzburg, Germany\\
$^4$Department of Physics, Technical University Dresden, 01069 Dresden, Germany\\}
\end{center}

\begin{abstract}
\end{abstract}
\maketitle

\vspace{-1em}
\tableofcontents
\vspace{1em}
\hrule
\vspace{1.5em}

\section{I. Derivation of the polarization term in the Extended Hubbard model}
Here, we give a detailed derivation of the effect of atomic screening on the one-dimensional extended Hubbard model (EHM) Hamiltonian, which was first derived in Ref. \cite{meinders1995atomic}. It plays an important role when dealing with real materials and in such cases it is necessary to include effective Coulomb repulsion $U_{eff}$, which is basically screened by atomic polarization. In most cases, this effective Coulomb repulsion, $U_{eff}$, is much smaller than bare Coulomb repulsion, $U$.  For instance, while Cu$^{2+}$ has \( U \approx 17 \) eV, the effective \( U \) for $Cu^{2+}$ in La$_2$CuO$_4$ is 10 eV \cite{annett1989electronic,hybertsen1990renormalization,grant1991realistic} and can even become negative for Bi$^{4+}$ in BaBiO$_3$\cite{varma1988missing}. To derive a Hamiltonian that incorporates the charge-dipole interaction, we consider a gapped system with an energy gap \( \Delta_{\alpha,\beta} \) between the lower orbital (\(\alpha\)) and higher orbital (\(\beta\)). Due to the electric field created by the electrons, a dipole moment $P_i$ at site $i$ is generated and the second quantized expression for it is given by,
\begin{equation}
    \bold{P_i}=-e\bold{R_{\alpha,\beta}}(c^{\dagger}_{i,\beta}c_{i,\alpha}+H.c.),
\end{equation}
where  $\bold{R_{\alpha,\beta}}=$$\int$$\Phi_{i,\alpha}$ $\bold{r}$ $\Phi_{i,\beta}$ d$^{3}$r. $\Phi_{i,\alpha(\beta)}$ corresponds to the real space wave function for the orbital $\alpha$ ($\beta$) of the $i$th site,  and $c^{\dagger}_{i,\alpha}$ ($c_{i,\alpha}$) is the creation (annihilation) operator for an electron at site $i$ and orbital $\alpha$. The expression for the electric field, $\bold{F_i}$ created by the neighboring charges in the system at a distance $a$ is
\begin{equation}
    \bold{F_{i}}={{e}\over{a^2}}\sum_{j}n_j\bm{\delta}_{ij} .
\end{equation}
Here, $\bm{\delta}_{ij}$ is the unit position vector between the neighboring sites $i$ and $j$. Since the dipoles are present in this electric field, the potential energy corresponding to that is given below:
\begin{equation}
    H_{pot}=-\sum_{i}\bold{P_{i}\cdot\bold{F_i}}
\end{equation}
So, the total Hamiltonian due to the polarization effect is given by,
\begin{equation}
H_{pol}=H_{pot}+\sum_{i}\Delta_{\alpha,\beta} c_{i,\beta}^{\dagger}c_{i,\beta}
\end{equation}
Substituting Eqs.(S1) and (S2), we can get the expression for $H_{pol}$ in one dimension is shown below

\begin{equation}
    H_{pol}={{e^2}\over{a^2}}|R_{\alpha,\beta}|\sum_{i}(n_{i+1}-n_{i-1})(c^{\dagger}_{i,\beta}c_{i,\alpha}+H.c.)+\Delta_{\alpha,\beta}c^{\dagger}_{i,\beta}c_{i,\beta}
\end{equation}

To derive the single-band model from this multi-band picture one need to use the perturbation theory and we can take the first term of $H_{pol}$ as a perturbation as this energy scale is much smaller compared to $\Delta_{\alpha,\beta}$. Considering the electron to be present in the $\alpha$ orbital and the higher lying orbital to be $\beta$, the zeroth order wave function is $|\Psi_{0}>$ and $|\Psi_i>$ is the excited state which can be generated from $|\Psi_0>$ using the expression $|\Psi_i>=c_{i,\beta}^{\dagger}c_{i,\alpha}|\Psi_0>$ . So, the zeroth order energies of the ground state and the excited state are $E_0$ and $E_0+\Delta_{\alpha,\beta}$, respectively. Now, the energies corresponding to the second-order correction give the form
\begin{equation}
    E=E_{0}-\sum_{i}{{|<\Psi_i|H_{pot}|\Psi_0>|^2}\over{\Delta_{\alpha,\beta}}}
\end{equation}

It simply give the expression $E=E_{0}-\sum_{i}{{(eR_{\alpha,\beta})^2}\over{\Delta_{\alpha,\beta}}}F_i^2$ and after putting the expression for $F_i$ we get the final form of the Hamiltonian:
\begin{equation}
    H_{pol}=-P\sum_{i}\left(\sum_{j\in {n.n}}n_j\bm{\delta_{ij}}\right)^2
\end{equation}
Here $P={{e^2R_{\alpha,\beta}^2}\over{a^4\Delta_{\alpha,\beta}}}$. So, the total Hamiltonian for the EHM with charge-dipole interaction is given by
\begin{equation}
    H_{tot}=H_{EHM}+H_{pol},
\end{equation}
where $H_{EHM}$ is the Hamiltonian for the EHM which includes the kinetic energy term, on-site Coulomb repulsion term, and nearest-neighbor Coulomb repulsion term of the electrons.

\newpage

\section{I\hspace{-1.2pt}I. Ground-state degeneracy at the atomic limit ($t=0$)}

\begin{figure}[h]
    \centering
    \includegraphics[width=0.8\linewidth]{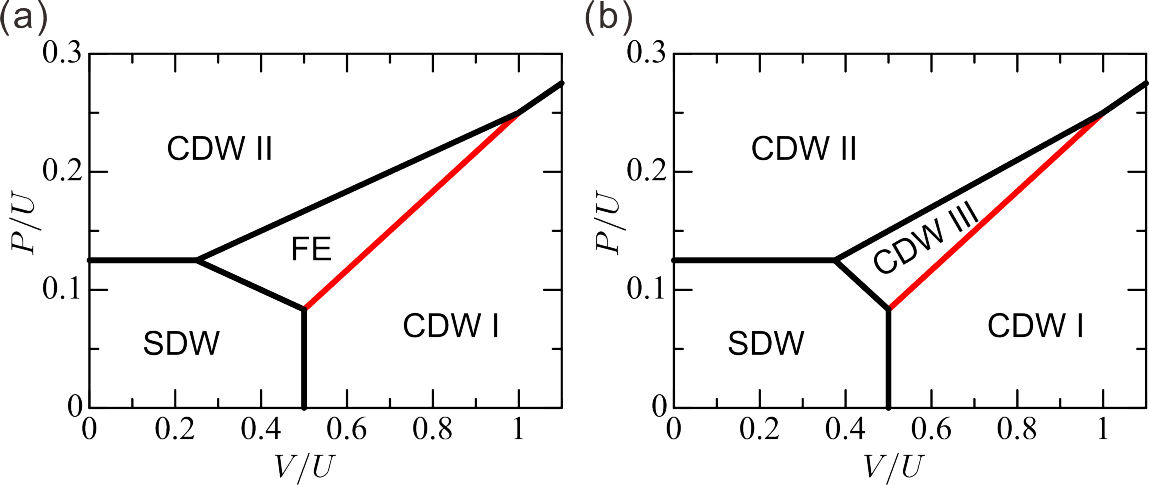}
    \caption{(a) Phase diagram of the system (1) in the atomic limit ($t=0$) within the ($V/U$, $P/U$) parameter space. The FE, CDW I, and CDW III states are energetically degenerate along the red line. (b) Phase diagram obtained by excluding the possibility of the FE phase from the analysis.}
    \label{fig:t0PD_SM}
\end{figure}

Our model, defined in Eq.~(1) of the main text, can be solved exactly in the atomic limit, $t=0$. In this limit, we consider five distinct phases characterized by energetically feasible charge distributions: spin-density wave (SDW), three types of charge-density wave (CDW I, CDW II, and CDW III), and a ferroelectric (FE) phase. The energy per site for each configuration is given by:
  \begin{subequations}
	\begin{align}
		\nonumber
		{\rm SDW}&: V+2P\\
		\nonumber
		{\rm CDW\ I}&: \frac{1}{2}U+2P\\
		\nonumber
		{\rm CDW\ II}&: \frac{1}{2}U+V-2P\\
		\nonumber
		{\rm CDW\ III}&: \frac{3}{8}U+\frac{1}{2}V+\frac{1}{2}P\\
		\nonumber
		{\rm FE}&: \frac{1}{3}U+\frac{2}{3}V\\
		\nonumber
	\end{align}
\end{subequations}

By comparing these energies and identifying the phase with the lowest energy at each point in parameter space, we obtain the phase diagram shown in Fig.~\ref{fig:t0PD_SM}(a), plotted in the $(V/U, P/U)$ plane. While the CDW III phase is not explicitly indicated in the diagram, its energy becomes degenerate along the boundary separating the FE and CDW I phases.

If we exclude the possibility of the FE phase from the analysis, the resulting phase diagram, shown in Fig.~\ref{fig:t0PD_SM}(b), reveals that regions previously occupied by the FE phase are replaced by the CDW III phase, particularly in the vicinity of the CDW I phase. This observation highlights the close competition between the FE and CDW III states in this region of parameter space.

Indeed, as discussed in the main text, introducing quantum fluctuations via finite hopping $t$ destabilizes the FE phase in favor of the CDW III phase, demonstrating the delicate balance between these competing orders in the atomic limit.

\newpage

\section{I\hspace{-1.2pt}I\hspace{-1.2pt}I. Demonstration of level-crossing method}
\begin{figure}[h!]
    \includegraphics[width=0.5\linewidth]{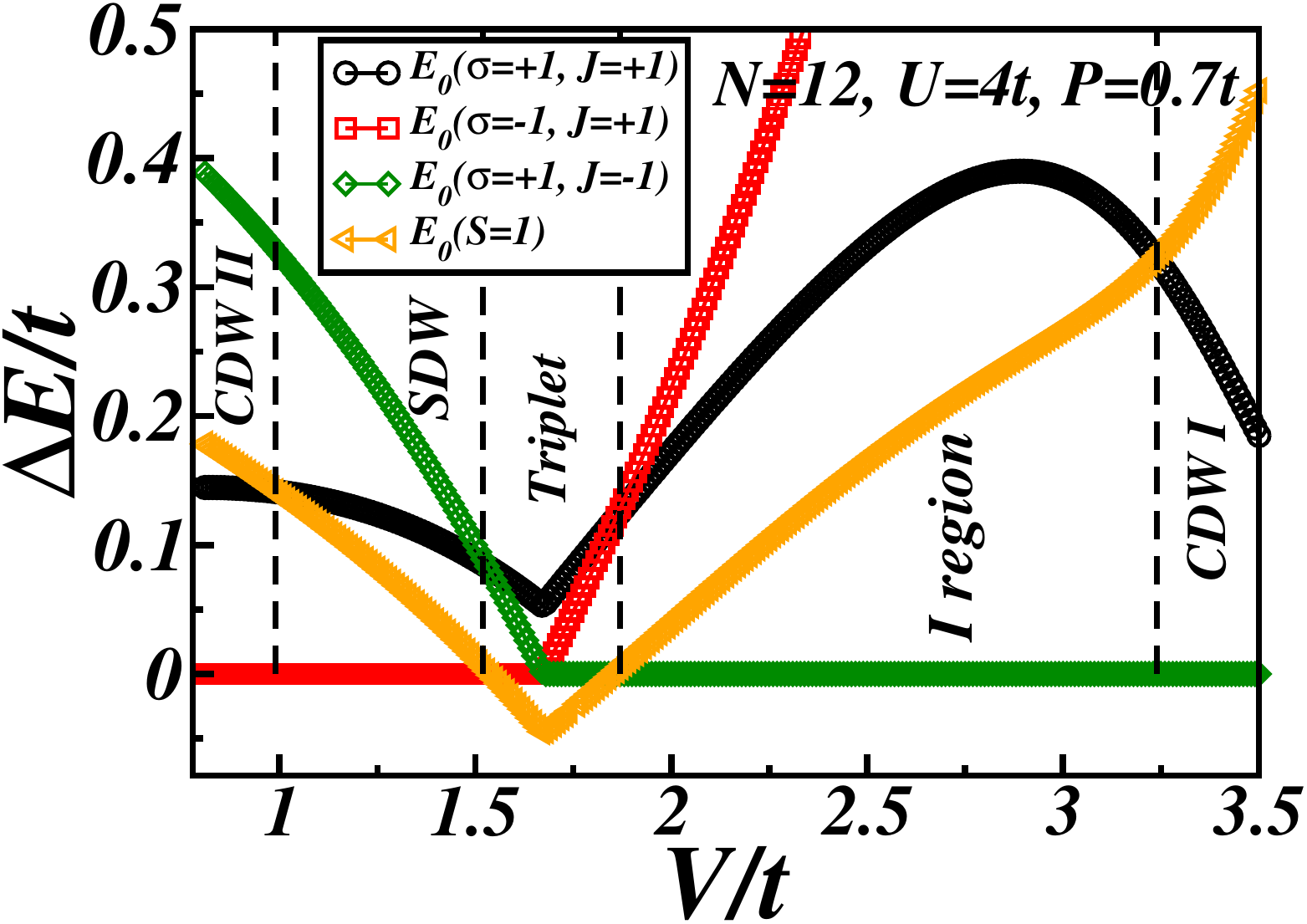}
    \caption{Energy gaps between inversion symmetry ($\sigma)$, e-h symmetry (J) and triplet ($S=1$) energy for $12$ site system using ED with PBC.  }
    \label{figS2}
\end{figure}

The Hamiltonian for the EHM with charge-dipole interaction, given in Eq. (1) of the main text, conserves total number of electrons ($N_e$), total spin of the system ($S$), and it's $z$-component. Consequently, at half-filling, it conserves electron-hole ($e$-$h$) symmetry ($J$ = $\pm$ 1), and the model on a ring also preserves the inversion symmetry ($C_i$), which is labeled as $\sigma = \pm 1$. The Lieb-Schultz-Mattis theorem states that the properties of the lowest-energy excitations can give information about the nature of the ground state (gs)\cite{LIEB1961407,oshikawa2000commensurability,nachtergaele2007multi} and the level crossing of excited states can be used to determine the phase boundary in a finite system \cite{nakamura2000tricritical,nakamura1999mechanism}. In Hubbard model at half-filling on a chain with periodic boundary condition shows various energies crossover and these crossing points are associated with symmetries crossover of the gs. In the thermodynamic limit, the lowest excited states either become degenerate or gapless with the gs. Therefore, if the excited states change, then the gs may also change. In this system if the gs is singlet and lowest excited state is triplet, then the gs shows a spin density wave due to breaking of the spin-parity symmetry. In case of I region also the lowest excited state is triplet. In a finite system if ground state and excited state is connected by inversion symmetry then gs show bond-order wave (BOW) phase as the degenerate gs breaks the inversion symmetry in the thermodynamic limit. Similarly, if the two lowest-energy states are related by electron-hole symmetry, the ground state spontaneously breaks this symmetry in the thermodynamic limit, resulting in a CDW phase. However, in the large-$U$ limit, CDW I and CDW II exhibit two-fold and four-fold degenerate ground states under periodic boundary conditions, respectively.

To identify the phase boundaries, we calculate energy crossovers of two lowest excited state using exact diagonalization (ED) for ring geometry in the valence bond basis \cite{ramasesha1984diagrammatic,soos1984valence,kumar2010bond}. We exploit the $e-h$ and $C_i$ symmetries and these crossover criterion work in most cases  except for some exotic cases. The results for $U = 4t$ and $P = 0.7t$ are presented in Fig. \ref{figS2}. By varying $V$ from $0.75t$ to $3.5t$, we observe three level crossings. In the intermediate region, the triplet state becomes the ground state. However, as the system size increases, this region diminishes, indicating that the triplet state arises due to finite-size effects. Below this region, at $V = 0.99034t$, a crossover occurs between the states $E_0(\sigma=+1, J=+1)$ and $E_0(S=1)$, marking the phase transition from the CDW II phase to the SDW phase where the gs is singlet and the first excited state is triplet. As $V$ increases further, another crossover is observed at $V = 3.23613t$, indicating a transition from region I to the CDW I phase, where the states $E_0(S=1)$ and $E_0(\sigma=+1, J=+1)$ intersect which is connected with gs by e-h symmetry . Additionally, the phase transition between the SDW phase and region I corresponds to a crossover between the states $E_0(\sigma=-1, J=+1)$ and $E_0(\sigma=+1, J=-1)$, which occurs at $V = 1.6731t$.

\newpage
\section{I\hspace{-1.2pt}V. Details of DMRG Technique}
Density Matrix Renormalization Group (DMRG) is a state-of-art numerical technique that is primarily used to compute the ground state and a few low-lying excited states in low-dimensional systems \cite{white1992density,schollwock2005density}. In this work, we employ DMRG to solve the many-body Hamiltonian for large system sizes, up to $ N = 240 $. While DMRG is most effective for open boundary conditions (OBC), boundary effects require careful treatment. To mitigate the effects of missing terms at open edge, we introduce correction fields of strength $V + 2P$ at the boundary sites ($ i = 1, N $) and $2P$ at the nearest-neighbor sites of the end ($i = 2, N-1$).

In the main text, we evaluate the presence or absence of long-range charge order in the CDW state using several order parameters. In a finite open system, translation symmetry can be explicitly broken, allowing charge ordering to be directly observed by examining the local charge density of a single symmetry-broken ground state. Typically, Friedel oscillations decay toward the center of the system with increasing system size. However, if the amplitude at the center remains finite in the thermodynamic limit, it signals the existence of long-range order. To reliably detect CDW order in our calculations, we apply appropriate magnetic and electric fields to explicitly select a symmetry-broken state.

To achieve high accuracy, we retain up to the 10,000 eigenstates in the reduced density matrix, keeping the maximum truncation error below $\sim 10^{-7}$ in the worst cases. When necessary, physical quantities are extrapolated to the $ m \to \infty $ limit to further improve precision. Depending on the observable, periodic boundary conditions (PBC) are also employed in our DMRG calculations --- particularly for determining excitation gaps, structure factors, and the central charge.

\newpage

\section{V. Order parameters for the CDW I and CDW II states}

\begin{figure}[h]
    \centering
    \includegraphics[width=0.57\linewidth]{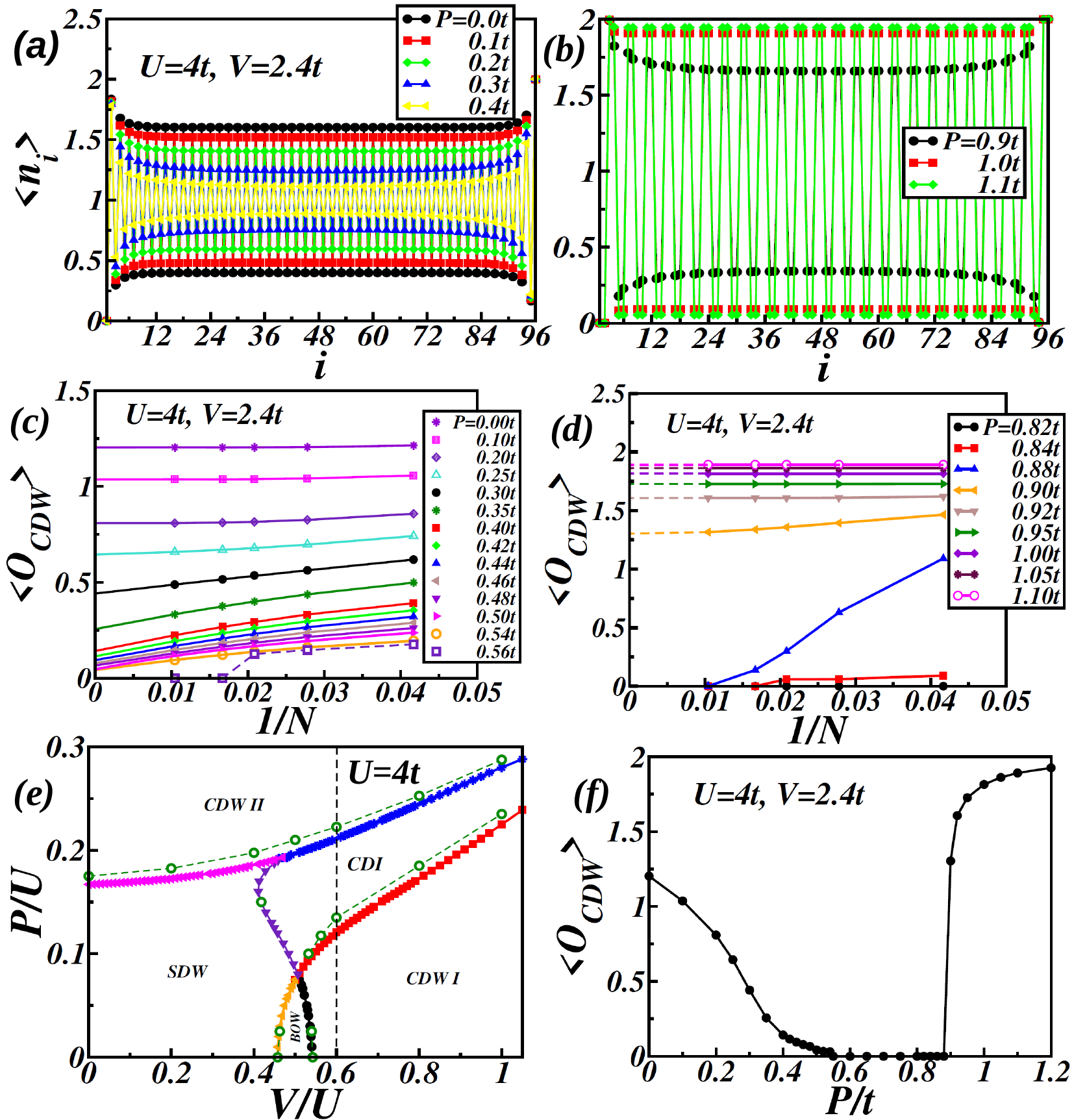}
    \caption{(a,b) Charge density distributions for representative parameter sets in the CDW I and CDW II phases, respectively, calculated on open chains with $N=96$ sites. (c,d) Finite-size scaling analyses for selected parameter sets in the CDW I and CDW II phases, respectively. (e) Ground-state phase diagram at $U=4t$, plotted as a function of $V/U$ and $P/U$. (f) Extrapolated order parameter $\langle {\cal O}_{\rm CDW} \rangle$ to the thermodynamic limit as a function of $P/t$ at $U=4t$ and $V=2.4t$. The scanned range corresponds to the dashed line shown in (e).}
    \label{fig:CDWOP_SM}
\end{figure}

This section explains how to detect the CDW I and CDW II states. Both are long-range charge-ordered phases, which can be directly observed by examining open clusters using the DMRG method. In finite systems with open boundary conditions, Friedel oscillations naturally break translational symmetry, allowing us to probe the charge order through the spatial distribution of the local charge density. Generally, these Friedel oscillations decay toward the center of the system as the system length increases. However, if the oscillation amplitude at the center remains finite regardless of system size, it signals the presence of long-range charge order. Thus, the order parameters for the CDW I and CDW II states are defined as $\langle {\cal O}_{\rm CDW} \rangle = |\langle n_{N/2} \rangle - \langle n_{N/2+1} \rangle|$. However, if the charge density distribution deviates from ``$\cdots$\hspace{0.2cm}$\uparrow\downarrow$\hspace{0.2cm}$\circ$\hspace{0.2cm}$\uparrow\downarrow$\hspace{0.2cm}$\circ$\hspace{0.2cm}$\uparrow\downarrow$\hspace{0.2cm}$\circ$\hspace{0.2cm}$\uparrow\downarrow$\hspace{0.2cm}$\circ$\hspace{0.2cm}$\cdots$'' for CDW I and ``$\cdots$\hspace{0.2cm}$\uparrow\downarrow$\hspace{0.2cm}$\uparrow\downarrow$\hspace{0.2cm}$\circ$\hspace{0.2cm}$\circ$\hspace{0.2cm}$\uparrow\downarrow$\hspace{0.2cm}$\uparrow\downarrow$\hspace{0.2cm}$\circ$\hspace{0.2cm}$\circ$\hspace{0.2cm}\hspace{0.2cm}$\cdots$'' for CDW II, $\langle {\cal O}_{\rm CDW} \rangle=0$ immediately.

Figures \ref{fig:CDWOP_SM}(a) and \ref{fig:CDWOP_SM}(b) show the charge density distributions of the CDW I and CDW II phases, respectively, calculated for an open chain with $96$ sites. It can be seen that the amplitude near the center of the system varies depending on the stability of the CDW states. We then perform finite-size scaling analysis of the order parameters to check for the presence of long-range charge order; the scalings for CDW I and CDW II phases are shown in Figs.~\ref{fig:CDWOP_SM}(c) and \ref{fig:CDWOP_SM}(d). When the CDW state is stable, the size dependence of the order parameter is relatively small, but as the CDW state approaches the phase boundary with another phase, the CDW state becomes unstable and the order parameter decreases rapidly as the size increases. The results of the order parameter extrapolated to the thermodynamic limit, calculated along the dashed line in Fig.~\ref{fig:CDWOP_SM}(e), are plotted. Both order parameters are zero in the intermediate CDI phase. Interestingly, the CDI-CDW I transition is of a second-order, whereas the CDI-CDW II transition is first-order like. This is due to the fact that the transition from the CDI to the CDW I state can be continuous because of the polarization of charges between adjacent sites, while the CDW II state cannot be realized by continuous charge transfer from the CDI state.

\newpage

\section{V\hspace{-1.2pt}I. Spin gap in the CDW I and CDW II phases}

\begin{figure}[h]
    \centering
    \includegraphics[width=0.8\linewidth]{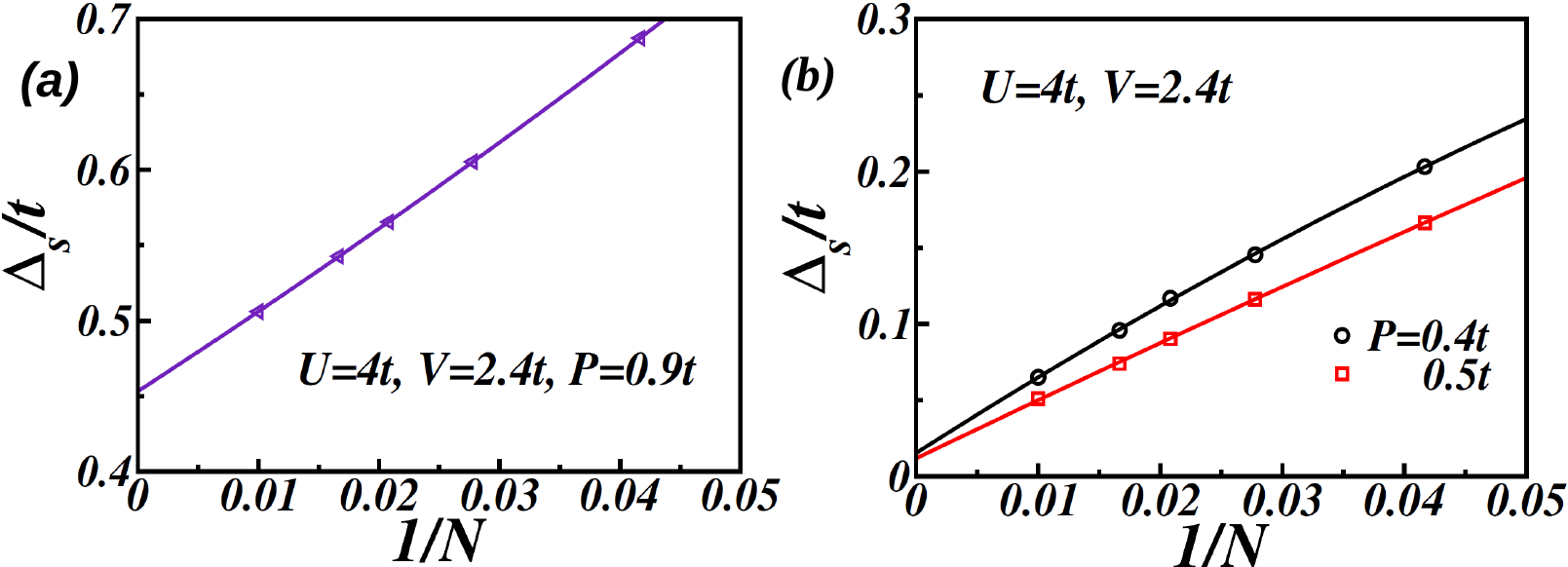}
    \caption{Finite-size scaling of the spin gap $\Delta_{\rm s}/t$ in (a) the CDW II phase and (b) the CDW I phase.
    }
    \label{fig:Ds_SM}
\end{figure}

The CDW I and CDW II phases appear in the ground-state phase diagram for any value of $U/t$. In broad terms, the CDW I phase emerges when $V$ is dominant, while the CDW II phase appears when $P$ is dominant. Since both phases consist of doubly occupied and empty sites, the spin gap is always open. Specifically, the spin gap in the CDW I phase approaches $\Delta_{\rm s}=3V-4P-U$ as $V$ increases, whereas in the CDW II phase, it approaches $\Delta_{\rm s}=12P-V-U$.

Even in parameter regions where the CDW phases are relatively unstable, DMRG calculations confirm the presence of a finite spin gap. Figures \ref{fig:Ds_SM}(a) and \ref{fig:Ds_SM}(b) show finite-size scaling of the spin gap in the CDW I and CDW II phases near the phase boundaries of the 
I region. Due to the instability of the CDW order in these regions, the spin gap exhibits relatively strong size dependence; nevertheless, the extrapolation indicates that the spin gap remains finite in the thermodynamic limit for all parameter values examined.

\newpage

\section{V\hspace{-1.2pt}I\hspace{-1.2pt}I. Properties of the SDW state}

\begin{figure}[h]
    \centering
    \includegraphics[width=0.8\linewidth]{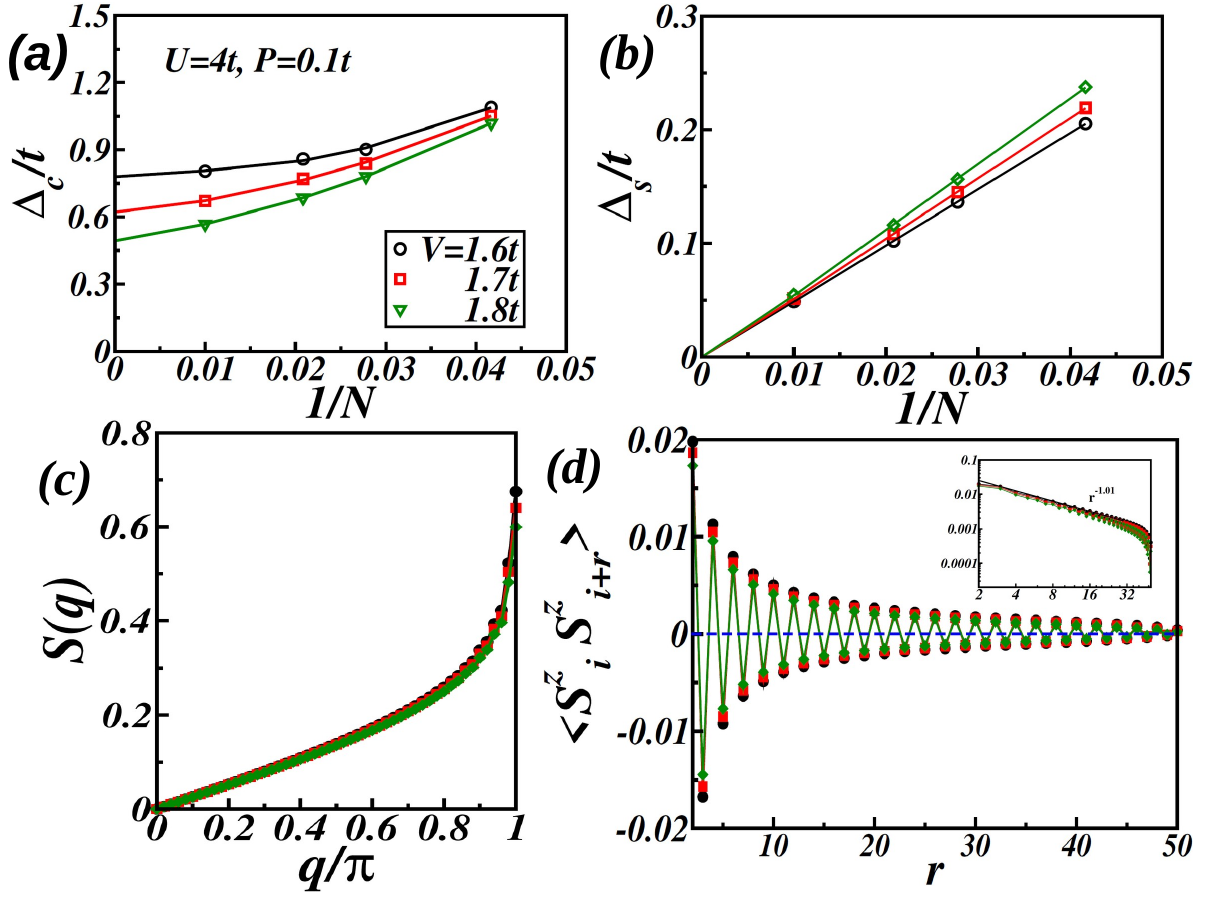}
    \caption{Results for three values of $V/t$ at $U=4t$ and $P=0.1t$, within the SDW phase. (a),(b) Finite-size scaling analyses for the charge gap and spin gap, respectively. (c) Spin structure factor $S(q)$. (d) Spin-spin correlation function as a function of distance. Inset: corresponding log-log plot.
    }
    \label{fig:SDW_SM}
\end{figure}

The SDW phase can be regarded as a Mott insulating phase. In this phase, the charge gap is open, while the spin gap remains closed, as shown in Figs.~\ref{fig:SDW_SM}(a) and \ref{fig:SDW_SM}(b). Regarding the spin degrees of freedom, the system can be effectively mapped onto an antiferromagnetic Heisenberg model. Consequently, the spin structure factor $S(q)$ exhibits a relatively sharp peak at $q=\pi$, as shown in Fig.~\ref{fig:SDW_SM}(c). However, as illustrated in Fig.~\ref{fig:SDW_SM}(d), the spin-spin correlation function decays with distance in a power-law manner, indicating the absence of true long-range magnetic order.

\newpage

\section{V\hspace{-1.2pt}I\hspace{-1.2pt}I\hspace{-1.2pt}I. Finite-size effects appearing near the SDW-CDI transition}

\begin{figure}[h]
    \centering
    \includegraphics[width=1.0\linewidth]{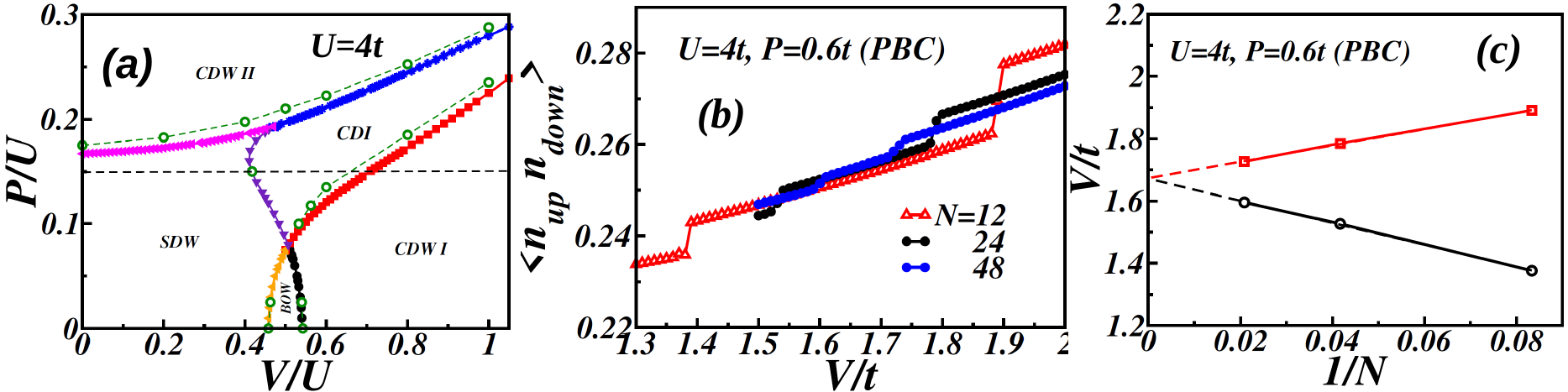}
    \caption{(a) Ground-state phase diagram at $U=4t$ in the ($V/U$,$P/U$) plane. (b) Double occupancy $\langle n_{i,\uparrow} n_{i,\downarrow} \rangle$ as a function of $V/t$, calculated for $N=12$, $24$, and $48$ at $U=4t$ and $P=0.6t$. (c) Finite-size scaling of the level-crossing points observed in (b).
    }
    \label{fig:S1_SM}
\end{figure}

Here, we comment on the finite-size effects observed in the periodic chain results near the SDW-CDI transition. In the level-crossing analysis described in the main text, the ground state is assumed to be a singlet ($S=0$). However, for finite-size periodic chains near the SDW-CDI transition, we find that the lowest-energy state is actually a triplet ($S=1$) state. Although this triplet state is expected to lie higher in energy than the singlet state in the thermodynamic limit, we examine this behavior in more detail.

Figure~\ref{fig:S1_SM}(a) shows the ground-state phase diagram at $U=4t$. Along the dashed line, we track the quantum numbers of the lowest-energy state obtained for the periodic chain. Figure~\ref{fig:S1_SM}(b) presents the double occupancy $\langle n_{i,\uparrow} n_{i,\downarrow} \rangle$ at $U=4t$ and $P=0.6t$ as a function of $V/t$, calculated for three different system sizes: $N=12$, $24$, and $48$. For each system size, two distinct jumps (level crossings) in the double occupancy are observed. The region between these jumps corresponds to the $S=1$ state, while the regions on either side correspond to the $S=0$ state.

As the system size increases, the width of the intermediate $S=1$ region becomes narrower. To quantify this, we extrapolate the positions of the two level-crossing points as a function of system size. The results are shown in Fig.~\ref{fig:S1_SM}(c). Interestingly, the two level-crossing points converge to a single point in the thermodynamic limit, at $V=1.67t$. This value coincides with the transition point between the SDW and CDI phases.

These results indicate that, in the thermodynamic limit, the $S=0$ and $S=1$ states become degenerate precisely at the SDW-CDI transition point. Outside this point, the $S=0$ state remains the true ground state. This finding supports the interpretation that the on-site effective Coulomb repulsion vanishes exactly at the SDW-CDI transition, as discussed in the main text.

\newpage

\section{I\hspace{-1.2pt}X. Additional central-charge analysis of the CDI phase}

\begin{figure}[h]
    \centering
    \includegraphics[width=0.8\linewidth]{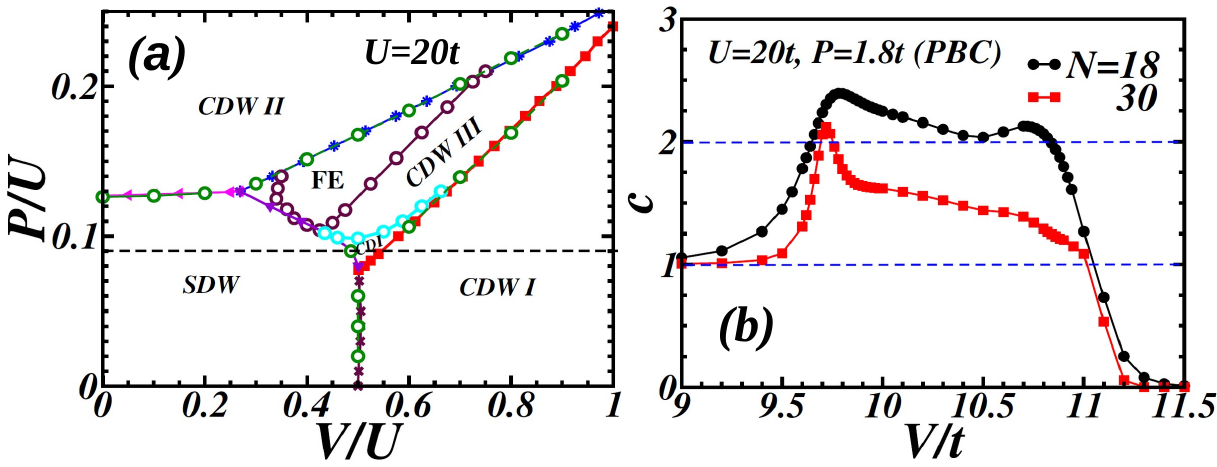}
    \caption{(a) Ground-state phase diagram at $U=20t$ in the ($V/U$,$P/U$) plane. (b) Central charge $c$ as a function of $V/t$ at $U=20t$ and $P=1.8t$, calculated for periodic chains with $N=18$ and $N=30$ sites.
    }
    \label{fig:cc_SM}
\end{figure}

In the main text, we demonstrated that the SDW and CDI transitions are continuous for relatively small $U/t$ (specifically, $U=4t$. To examine whether this behavior persists at larger $U/t$, we analyze several phase transitions using the central charge at $U=20t$. Figure \ref{fig:cc_SM}(a) shows the ground-state phase diagram in the ($V/U$, $P/U$) plane. In Fig.~\ref{fig:cc_SM}(b), we plot the central charge $c$, computed from periodic chains with $N=18$ and $N=30$ sites, as a function of $V/t$ at fixed $P=1.8t$.

The SDW-CDI transition occurs at $V=9.7t$, and the CDI-CDW I transition at $V=11.0t$. In both the SDW and CDI phases, the charge gap is finite while the spin gap is closed, resulting in one gapless mode ($c=1$). However, at the SDW-CDI transition point, we observe $c=2$, consistent with the behavior found at $U=4t$, indicating that the charge gap closes at this transition. In contrast, within the CDW I phase, both the charge and spin gaps are open, and consequently, the central charge drops to $c=0$, reflecting the absence of gapless excitations. Notably, at the CDI-CDW I transition, $c$ drops from 1 to 0, signaling that the spin gap opens upon entering the CDW I phase, while the charge gap remains finite throughout the transition.

\newpage

\section{X. Absence of charge and spin orders in the CDI state}

\begin{figure}[h]
    \centering
    \includegraphics[width=0.9\linewidth]{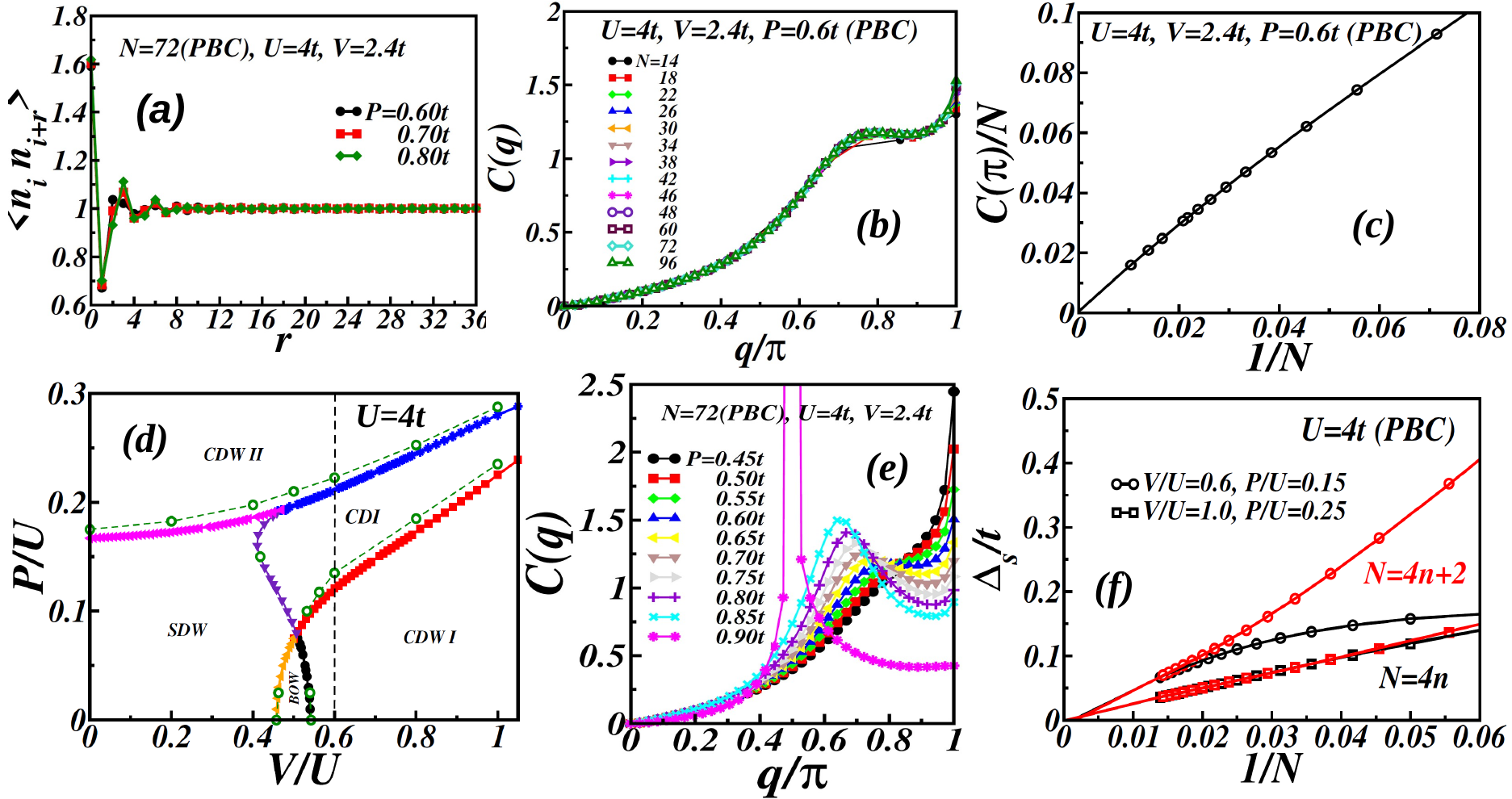}
    \caption{(a) Real-space density-density correlation $\langle n_i n_{i+r}\rangle$ as a function of distance for a periodic chain with $N=72$, $U=4t, V=2.4t$.  
    (b) Charge structure factor $C(q)$ at $U=4t, V=2.4t,$ and $P=0.6t$, calculated for different system sizes using PBC.
    (c) Finite-size scaling of \( C(q = \pi)/N \).
    (d) Phase diagram at $U=4t$ in the ($V/U$, $P/U$) plane. (e) Evolution of \( C(q) \) with increasing polarization \( P \) from $0.45t$ to $0.90t$ for $N=72$, $U=4t$ and $V=2.4t$ using PBC. (e) Finite size scaling of the spin gap $\Delta_s/t$ in the CDI phase.
}
    \label{fig:CDI_SM}
\end{figure}

As discussed in the main text, although the charge gap is open in the CDI phase, no charge polarization is observed. To further confirm this, Fig.~\ref{fig:CDI_SM}(a) shows the distance dependence of the density-density correlation function for several parameter sets within the CDI phase. While short-range density modulations are present, they decay rapidly with distance and converge to $\langle n_in_j \rangle \sim 1$, indicating the absence of long-range charge order.

Additionally, the charge structure factor, $\tilde{C}(q)$ at \( U = 4t \), \( V = 2.4t \), and \( P = 0.6t \) exhibits a sharp, delta-function-like peak at \( q = 0 \), as shown in the main text. A secondary peak also appears at \( q = \pi \), which is more clearly visible in the plot of \( C(q) \) in Fig. \ref{fig:CDI_SM}(b). However, this secondary peak does not correspond to long-range charge order. To verify the absence of long-range charge order, we extrapolate \( C(q=\pi)/N \) to the thermodynamic limit, where it vanishes, as demonstrated in Fig. \ref{fig:CDI_SM}(c).
The evolution of $C(q)$ as $P$ increases is shown in Fig.~\ref{fig:CDI_SM}(e) 
along the dotted line shown in the phase diagram in Fig. \ref{fig:CDI_SM}(d).
At $P = 0.6t$, there is a shoulder feature near $q = 0.8\pi$, which develops and shifts to lower $q$ as $P$ increases. Nevertheless, this behavior does not indicate the onset of charge order. In contrast, upon crossing the transition into the CDW II phase (around $P \sim 0.9t$), a sharp peak emerges at $q = \pi/2$, signaling the establishment of long-range charge order characteristic of the CDW II phase.

Thus, the CDI phase exhibits no charge order and remains insulating, despite the absence of charge ordering and the presence of an effectively attractive on-site interaction. This constitutes a highly unconventional insulating state. A more detailed discussion of the mechanisms responsible for this unexpected insulating behavior is provided in the following sections.

Additionally, we evaluate the spin gap to probe the magnetic sector. Fig.~\ref{fig:CDI_SM}(f) shows finite-size scaling of $\Delta_{\rm s}$ for representative points within the CDI region. In both cases, $\Delta_{\rm s}/t$ extrapolates to zero, indicating gapless spin excitations.

\newpage

\section{X\hspace{-1.2pt}I. Mixed short-range charge orders in CDI states}

\begin{figure}[h]
    \centering
    \includegraphics[width=1.0\linewidth]{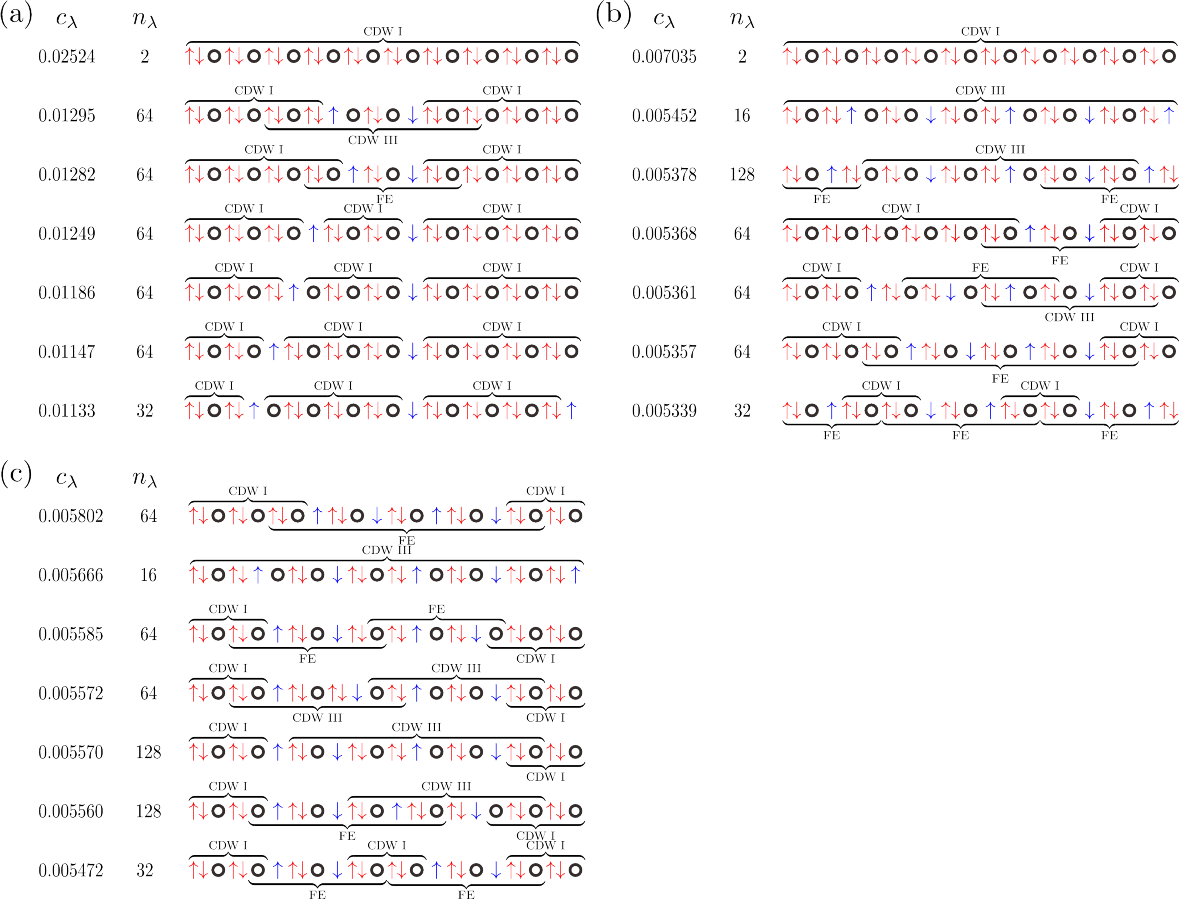}
    \caption{Seven dominant configurations with the largest coefficients in the ground-state wave function obtained from ED calculations on a 16-site system for parameters: (a) $U=4t$, $V=3.2t$, $P=0.8t$, (b) $U=4t$, $V=2.4t$, $P=0.7t$, and (c) $U=4t$, $V=2.4t$, $P=0.8t$. Here, $c_\lambda$ denotes the coefficient value of each configuration, and $n_\lambda$ indicates the number of physically equivalent configurations.}
    \label{fig:coeff_SM}
\end{figure}

The CDI state exhibits unique properties. The system remains insulating even though the effective on-site Coulomb repulsion is negative and there is no long-range charge ordering. To understand the origin of this insulating behavior, we examine the ground-state wavefunctions in the CDI phase. Figure~\ref{fig:coeff_SM} shows the electron configurations with the largest weights in the ground-state wavefunction, obtained from ED calculations on a 16-site periodic chain for three parameter sets: (a) $U=4t$, $V=3.2t$, $P=0.8t$, (b) $U=4t$, $V=2.4t$, $P=0.7t$, and (c) $U=4t$, $V=2.4t$, $P=0.8t$. 

For the case with $U=4t$, $V=3.2t$, and $P=0.8t$ [Fig.~\ref{fig:coeff_SM}(a)], the configuration with the largest weight corresponds to a CDW I pattern. However, this does not develop into a long-range order. The second most significant configuration represents a mixture of CDW I and CDW III states, while the third combines CDW I with FE features. Subsequent configurations involve electrons trapped between regions exhibiting short-range CDW I configuration. These local structures restrict the mobility of charge carriers, severely hindering electron propagation and resulting in an insulating state.

For the second parameter set, $U=4t$, $V=2.4t$, and $P=0.7t$ [Fig.~\ref{fig:coeff_SM}(b)], the largest weight corresponds to a CDW I-type order, but again, we have confirmed that there is no indication of long-range order. It is quite interesting that a CDW III-type order has the second largest weight. This indeed indicates that the system is in a strong charge frustration. All subsequent configurations involve various combinations of short-range charge orders. As in the previous case, this mixture of competing local orders prevents coherent electron motion, leading to insulating behavior.

For the third parameter set, $U=4t$, $V=2.4t$, and $P=0.8t$ [Fig.~\ref{fig:coeff_SM}(c)], none of the seven dominant configurations correspond to a pure CDW state. The configuration associated with the CDW I phase appears only at the 15th position ($n_\lambda=2$, $c_\lambda=0.004093$). This provides further evidence of strong charge frustration. As in the previous two cases, the dominant configurations consist of various short-range CDW segments. All the coefficients in these parameters have very small values due to high quantum fluctuation in this phase  and correspond to the disordered phase.

In summary, the ground-state wavefunction in the CDI phase is predominantly composed of patterns with short-range charge ordering, reminiscent of nearby charge-ordered phases. This local ordering results in a high density of double occupancy regions that act as barriers to charge transport, effectively localizing the electrons and stabilizing an insulating state despite the absence of long-range order and negative on-site Coulomb interaction.

\newpage
\section{X\hspace{-1.2pt}I\hspace{-1.2pt}I. Confirmation of insulating behavior in the CDI phase via single-particle spectral analysis}

\begin{figure}[h]
    \centering
    \includegraphics[width=0.71\linewidth]{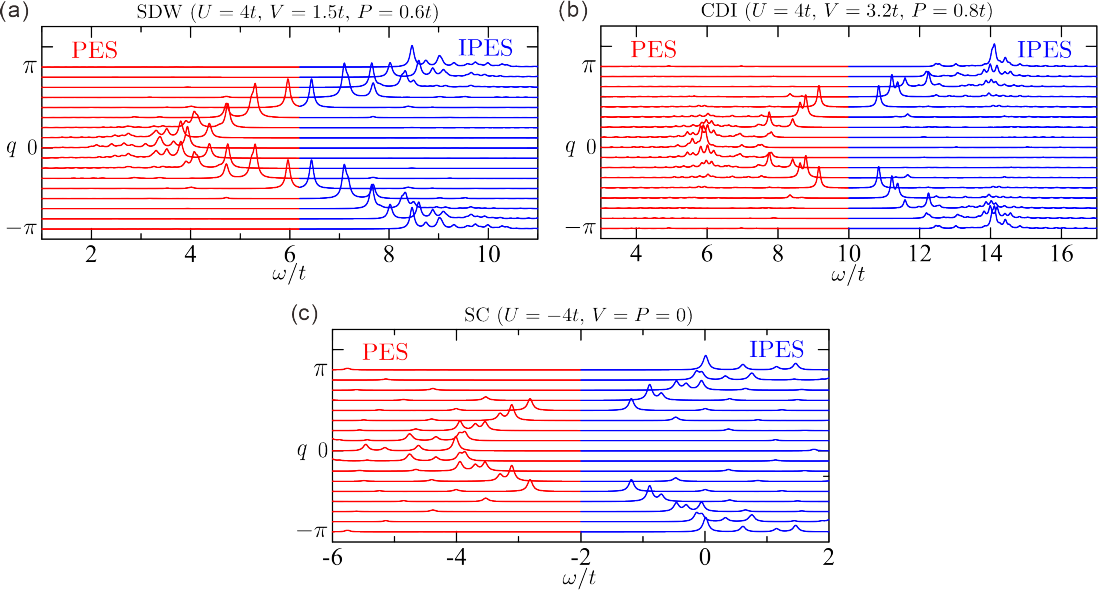}
    \caption{Single-particle excitation spectra calculated using the 16-site chain for (a) SDW, (b) CDI, and (c) SC states.}
    \label{fig:PESBIS_SM}
\end{figure}

As described above, the CDI phase lacks long-range charge order and instead consists of a mixture of several short-range charge configurations. This results in an insulating state due to the inhibition of coherent electron propagation. To confirm the nature of the electronic dispersion in the CDI state, we calculate the single-particle excitation spectrum, defined as
\begin{eqnarray}
	A(q,\omega) = A^-(q,-\omega) + A^+(q,\omega)
	\label{Aqw}
\end{eqnarray}
where $A^-(q,\omega)$ is the photoemission (PES) spectrum,
\begin{eqnarray}
	A^-(q,\omega) = \sum_{\nu \sigma} |\langle \psi^{N_e-1}_\nu |c_{q\sigma}| \psi^{N_e}_0 \rangle|^2 \delta(\omega-E^{N_e-1}_\nu+E^{N_e}_0)
	\label{PES}
\end{eqnarray}
and $A^+(q,\omega)$ is the inverse photoemission (IPES) spectrum,
\begin{eqnarray}
	A^+(q,\omega) = \sum_{\nu \sigma} |\langle \psi^{N_e+1}_\nu |c^\dagger_{q\sigma}| \psi^{N_e}_0 \rangle|^2 \delta(\omega-E^{N_e+1}_\nu+E^{N_e}_0).
	\label{IPES}
\end{eqnarray}
Here, $| \psi^{N_e}_\nu \rangle$ and $E^{N_e}_\nu$ denote the $\nu$-th eigenstate and eigenenergy of the system with $N_e$ electrons, where $\nu=0$ corresponds to the ground state. The operators $c^\dagger_{q\sigma}$ ($c_{q\sigma}$) are the Fourier transforms of the real-space creation (annihilation) operators $c^\dagger_{i\sigma}$ ($c_{i\sigma}$). Specifically,
\begin{eqnarray}
	c_{q\sigma}=\frac{1}{\sqrt{N}}\sum_i 
	e^{iqr_i}c_{i\sigma}. 
	\label{cFT}
\end{eqnarray}

We perform the calculations using the Lanczos method under periodic boundary conditions for a 16-site chain. Figures~\ref{fig:PESBIS_SM}(a) and (b) show the calculated $A(q,\omega)$ for representative parameters corresponding to the SDW and CDI phases, respectively. In the SDW phase, a well-defined cosine-like band dispersion is clearly visible, although the system is in a Mott insulating state.

In contrast, the CDI phase exhibits an ill-defined dispersion in addition to the opening of a single-particle charge gap. The absence of a coherent band structure indicates a suppression of electron mobility and the breakdown of quasiparticle propagation. This behavior is consistent with a localized, insulating state, where charge carriers are unable to move freely through the system. In particular, the high density of double-occupancy sites in the CDI phase effectively acts as scattering centers or barriers that obstruct electron motion, reinforcing the insulating character.

For comparison, we also examine $A(q,\omega)$ for the negative-$U$ Hubbard chain, which features a similarly high density of double-occupancy sites. The results are presented in Fig.~\ref{fig:PESBIS_SM}(c). Despite the large number of double occupancies, the cosine-like dispersion remains sharp and well-defined in this case. This reflects the fact that the ground state belongs to a superconducting (SC) phase. While a single-particle gap is open, the two-particle gap is closed, and the system exhibits metallic behavior due to coherent pair transport.

\newpage

\section{X\hspace{-1.2pt}I\hspace{-1.2pt}I\hspace{-1.2pt}I. Mapping of FE and CDW III states onto effective spin models}

\begin{figure}[h!]
    \centering
    \includegraphics[width=1.0\linewidth]{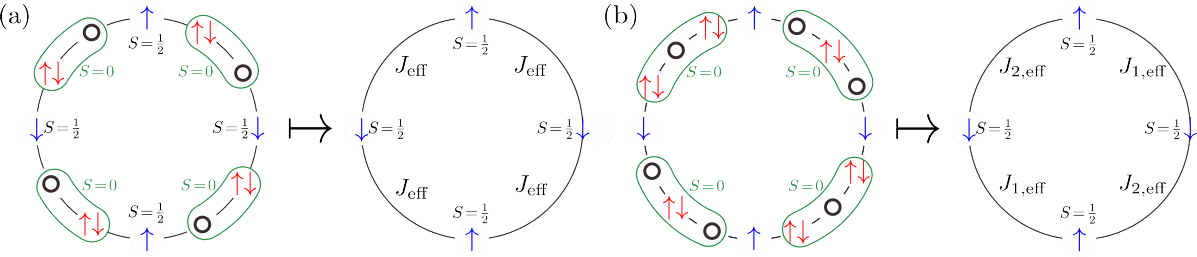}
    \caption{Small clusters employed for the effective spin-model mapping. (a) Original 12-site chain illustrating the FE state and its corresponding 4-site Heisenberg spin chain representation. (b) Original 16-site chain displaying the CDW III and its corresponding 4-site spin-Peierls Heisenberg chain analog.}
    \label{fig:mapHeis}
\end{figure}

To gain deeper insights into the magnetic properties of the FE and CDW III states, we map their charge configurations onto effective spin models.

We first consider the FE state [Fig.~\ref{fig:mapHeis}(a)]. In this mapping, doubly occupied and empty sites are regarded as spin $S=0$, and singly occupied sites as spin $S=1/2$. Establishing a one-to-one correspondence between the original electronic model and an effective Heisenberg spin model allows us to estimate the effective exchange interaction. Specifically, the original 12-site chain can be effectively mapped onto a 4-site Heisenberg chain, as illustrated in Fig.~\ref{fig:mapHeis}(a). For the parameter set $U=20t$, $V=12.5t$, and $P=3.6t$, the original 12-site chain has a ground state ($S=0$) energy of $E_0(S=0)/t=178.40871374652$, while the lowest spin excitation ($S=1$) is at energy $E_0(S=1)/t=178.40965801554$, resulting in an energy difference $\Delta E_0(S=1)/t=0.000944269$. For a 4-site Heisenberg chain with exchange interaction $J_{\rm eff}$, we have $E_0(S=0)=-2J_{\rm eff}$ and $E_0(S=1)=-J_{\rm eff}$, giving $\Delta E_0(S=1)=J_{\rm eff}$. Matching the energy differences between the two systems, we obtain an effective exchange interaction $J_{\rm eff}/t=0.000944269$ for the FE state.

A similar approach can be applied to the CDW III state. However, as shown in Fig.~\ref{fig:mapHeis}(b), two distinct charge configurations-``$\uparrow\downarrow$ $\circ$ $\uparrow\downarrow$'' and ``$\circ$ $\uparrow\downarrow$ $\circ$''-alternate between singly occupied ($S=1/2$) sites. Thus, the mapped Heisenberg chain is expected to be of spin-Peierls type. To estimate the effective interactions, we compare ground and two excited state energies between the original 16-site system and the mapped 4-site spin-Peierls Heisenberg chain [Fig.~\ref{fig:mapHeis}(b)].

Considering parameters $U=20t$, $V=10.5t$, and $P=2.15t$ for the CDW III state, the original 16-site system energies are $E_0(S=0)/t=212.92566929516$, $E_0(S=1)/t=212.94088241681$, and $E_0(S=2)/t=212.97114671580$, giving energy differences $\Delta E_0(S=1)/t=0.015213122$ and $\Delta E_0(S=2)/t=0.045477421$. In the 4-site spin-Peierls chain with effective interactions $J_{\rm 1,eff}$ and $J_{\rm 2,eff}$, achieving the ratio of these energy differences requires setting $J_{\rm 2,eff}/J_{\rm 1,eff}=0.887124$. This yields energies $E_0(S=0)=-1.8921741395892J_{\rm 1,eff}$, $E_0(S=1)=-0.943562J_{\rm 1,eff}$, and $E_0(S=2)=0.943562J_{\rm 1,eff}$, that is, $\Delta E_0(S=1)=0.94861214J_{\rm 1,eff}$ and $\Delta E_0(S=2)=2.83573614J_{\rm 1,eff}$. This results in effective interactions $J_{\rm 1,eff}=0.016037241t$ and $J_{\rm 2,eff}=0.014227022t$. The spin gap obtained from this spin-Peierls configuration is $\Delta_{\rm s}=0.226715523J_{\rm 1,eff}$, corresponding to $\Delta_{\rm s}/t=0.0036358915$, consistent with the main text's value of $\Delta_{\rm s}/t \sim 0.0019$.

Investigating another CDW III regime with parameters $U=10t$, $V=8t$, and $P=2t$, we find energies for the original 16-site system: $E_0(S=0)/t=132.54016031518$, $E_0(S=1)/t=132.55017939290$, and $E_0(S=2)/t=132.57021447191$, giving $\Delta E_0(S=1)/t=0.010019078$ and $\Delta E_0(S=2)/t=0.030054157$. Matching these ratios in the spin-Peierls model requires $J_{\rm 2,eff}/J_{\rm 1,eff}=0.98$. Corresponding energies for the effective 4-site system are $E_0(S=0)=-1.9801515035589J_{\rm 1,eff}$, $E_0(S=1)=-0.99J_{\rm 1,eff}$, and $E_0(S=2)=0.99J_{\rm 1,eff}$. This yields $J_{\rm 1,eff}=0.010118732t$ and $J_{\rm 2,eff}=0.009916357t$, resulting in a significantly smaller spin gap $\Delta_{\rm s}/t=0.0005312495$, in good agreement with the main text's value of $\Delta_{\rm s}/t \sim 0.0003$.

These analyses reveal that decreased stability of the CDW III phase correlates with diminished spin-Peierls intensity. However, the detailed dependence of this effect on $U$ is complex, as the magnitude of effective interactions strongly depends on spin fluctuations.

\newpage
\section{X\hspace{-1.2pt}I\hspace{-1.2pt}V. Order parameter for the BOW phase}
\begin{figure}[h!]
    \centering
    \includegraphics[width=1.0\linewidth]{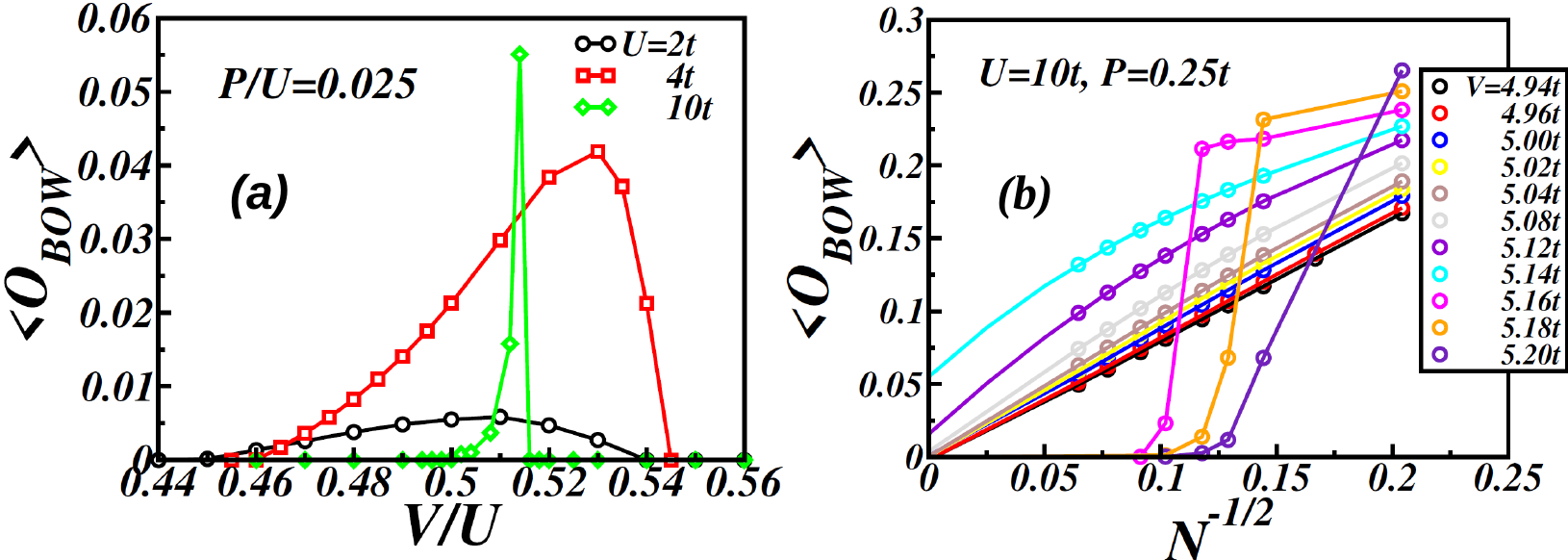}
    \caption{(a) Extrapolated values of the BOW order parameter for different values of $U$. (b) Finite-size scaling of the BOW order parameter for $U=10t$ and $P=0.25t$.}
    \label{S12}
\end{figure}

In this section, we present how the BOW state is detected using DMRG. Since the BOW state breaks translational symmetry, its order parameter can be defined as a local quantity under OBC. Specifically, the BOW order parameter is defined as  
\begin{equation}
    \langle {\cal O}_{\rm BOW} \rangle = \lim_{N \to \infty} \Big| \langle B_{N/2} - B_{N/2+1} \rangle \Big|,
\end{equation}
where $B_i$ is the BOW operator between sites $i$ and $i+1$, given by:
\begin{equation}
    B_i = \frac{1}{2} \sum_{\sigma} \left( c_{i,\sigma}^{\dagger} c_{i+1,\sigma} + H.c. \right).
\end{equation}
A nonzero amplitude of the oscillation of the BOW operator at the center of the system (\(\langle{\cal O}_{\rm BOW}\rangle \neq 0\)) signifies the long range ordered BOW phase.

Figure~\ref{S12}(a) presents the extrapolated values of $\langle {\cal O}_{\rm BOW} \rangle$ for different interaction strengths ($U = 2t, 4t,$ and $10t$), as a function of $V/U$, while keeping the ratio $P/U = 0.025$ fixed.  The results demonstrate that the BOW phase remains robust even in the presence of finite polarization $P$. For large $U/t$, the BOW phase is sandwiched between the SDW phase (on the small $P/U$ side) and the CDW I phase (on the large $P/U$ side), and the transitions become sharper with increasing $U/t$. These trends closely resemble the $P=0$ case.

Figure~\ref{S12}(b) presents the finite-size scaling of $\langle {\cal O}_{\rm BOW} \rangle$ for various values of $V/t$ near the SDW-BOW-CDW I phase boundaries at $U=10t$ and $P/U=0.025$. For $4.94 \le V/t \le 5.14$, the order parameter extrapolates smoothly to a finite value in the thermodynamic limit, indicating a continuous transition from the SDW to the BOW phase. In contrast, for $5.16 \le V/t \le 5.20$, the order parameter abruptly vanishes beyond a certain system size, suggesting that long-range BOW order is no longer sustained. This behavior is characteristic of a first-order transition - in this case, between the BOW and CDW I phases.

\newpage

\section{X\hspace{-1.2pt}V. Additional information for the FE phase}

\begin{figure}[h!]
    \centering
    \includegraphics[width=0.8\linewidth]{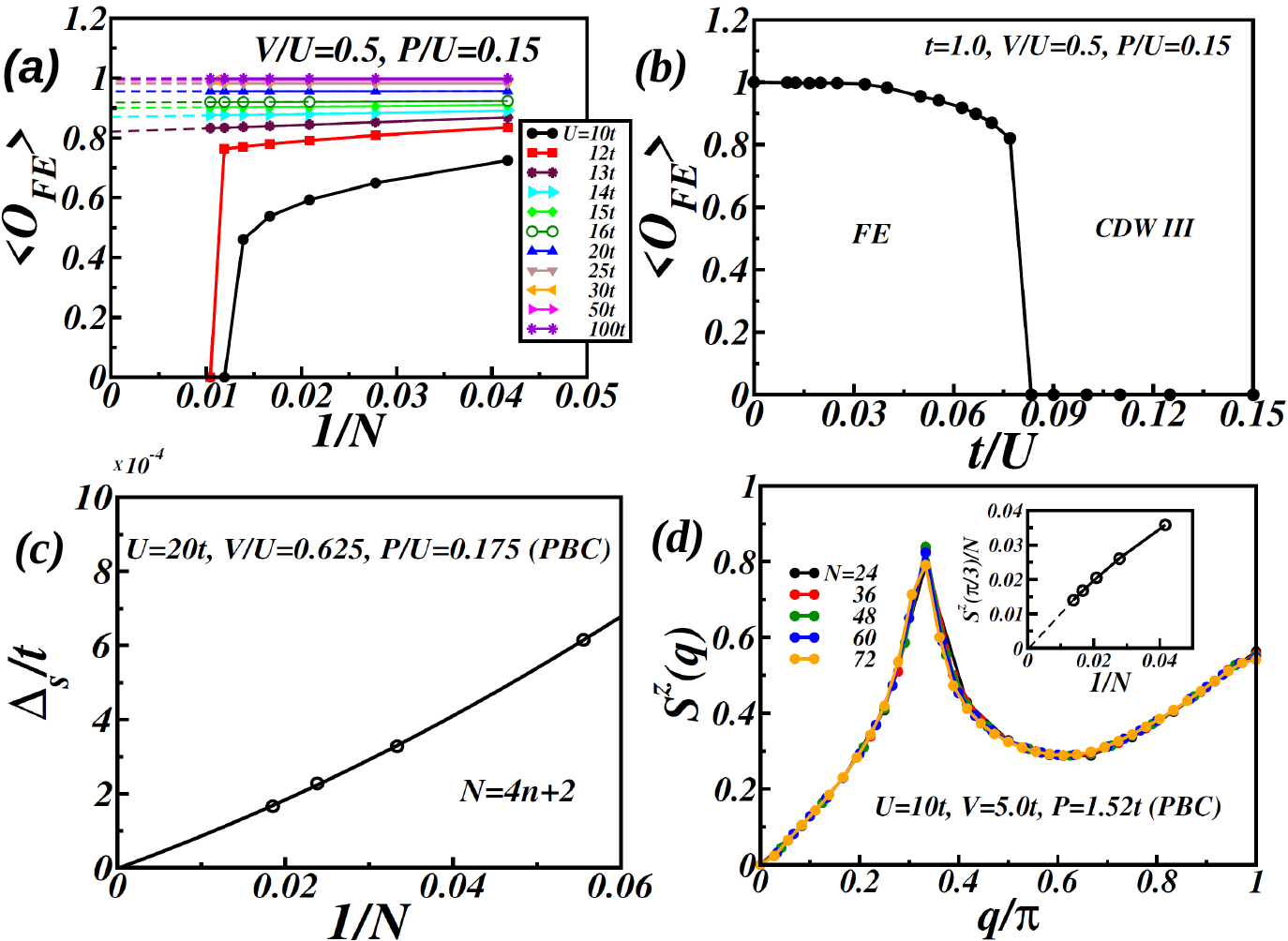}
    \caption{(a) Finite-size scaling of the FE order parameter, \(\langle {\cal O}_{\rm FE} \rangle\), for \(V/U = 0.5\) and \(P/U = 0.15\). (b) Extrapolated FE order parameter as a function of \(t/U\). (c) Finite-size scaling of the spin gap for a representative point in the FE phase. (d) Spin structure factor \(S^z(q)\) for \(U = 10t\), \(V = 5.0t\), and \(P = 1.52t\), computed using PBC for system sizes \(N = 24, 36, 48, 60,\) and \(72\). The inset shows the finite-size scaling of the peak height of \(S^z(q)\).
    }
    \label{S13}
\end{figure}

Figure~\ref{S13}(a) shows the finite-size scaling of the FE order parameter, \(\langle {\cal O}_{\rm FE} \rangle\), for fixed interaction ratios \(V/U = 0.5\) and \(P/U = 0.15\). The corresponding extrapolated values of \(\langle {\cal O}_{\rm FE} \rangle\) as a function of \(t/U\) are shown in Fig.~\ref{S13}(b). A finite FE order parameter in the thermodynamic limit at small \(t/U\) confirms the presence of long-range charge order in this phase. The transition between the FE and CDW III phases is found to be of first order.

As discussed in the main text and in Sec.~X\hspace{-1.2pt}I\hspace{-1.2pt}I\hspace{-1.2pt}I, the low-energy spin sector in the FE phase can be effectively mapped onto a uniform antiferromagnetic Heisenberg chain. In agreement with this mapping, the spin gap extrapolates to zero, as shown in Fig.~\ref{S13}(c), indicating gapless spin excitations.

To further characterize the spin sector, we compute the spin structure factor \(S^z(q)\), defined by
\begin{equation}
	S^z(q) = \sum_{r=0}^{N-1} \langle S_i^z S_{i+r}^z \rangle e^{iqr}.
\end{equation}
Figure~\ref{S13}(d) displays \(S^z(q)\) for \(U = 10t\), \(V = 5.0t\), and \(P = 1.52t\), calculated under PBC for system sizes \(N = 24, 36, 48, 60,\) and \(72\). All curves exhibit a peak at \(q = \pi/3\), indicating a spin periodicity of six in the FE phase. The extrapolated peak height approaches zero with increasing system size (inset), consistent with quasi-long-range spin correlations and critical spin behavior expected from the effective Heisenberg model.

\newpage

\section{X\hspace{-1.2pt}V\hspace{-1.2pt}I. Additional information for the CDW III phase}

\begin{figure}[h!]
	\centering
	\includegraphics[width=1.0\linewidth]{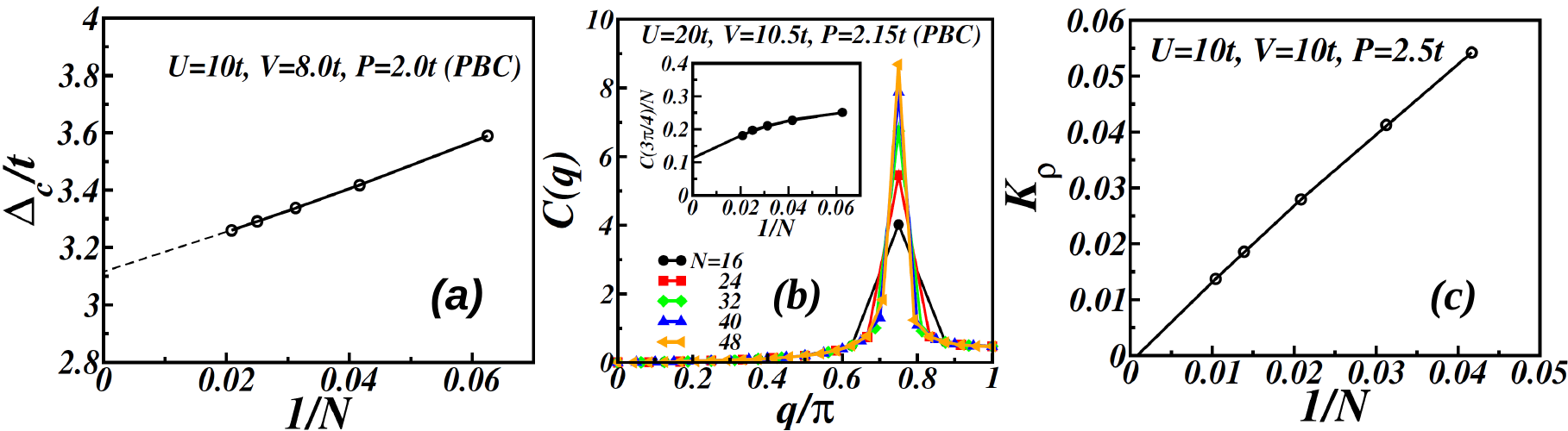}
	\caption{(a) Finite-size scaling of the charge gap for \( U = 10t \), \( V = 8.0t \), and \( P = 2.0t \), calculated using PBC. (b) Charge structure factor \( C(q) \) for \( U = 20t \), \( V = 10.5t \), and \( P = 2.15t \), computed using PBC. The inset shows the extrapolation of the peak height at \( q = 3\pi/4 \) to the thermodynamic limit. (c) Finite-size scaling of the Tomonaga-Luttinger liquid parameter \( K_\rho \) for \( U = 10t \), \( V = 10t \), and \( P = 2.5t \).}
	\label{S14}
\end{figure}

In this section, we provide further details on the CDW III phase. Since the charge gap should be finite in a charge-ordered phase, we calculate the charge gap for a representative parameter set for the CDW III phase. Figure~\ref{S14}(a) shows the finite-size scaling of the charge gap for \( U = 10t \), \( V = 8.0t \), and \( P = 2.0t \), calculated under PBC. The extrapolated value clearly remains finite in the thermodynamic limit, confirming the insulating nature of this phase.

Additionally, the charge structure factor \( C(q) \), shown in Fig.~\ref{S14}(b) for \( U = 20t \), \( V = 10.5t \), and \( P = 2.15t \), exhibits a pronounced peak at \( q = 3\pi/4 \), signaling a commensurate charge-density modulation. The inset displays the finite-size scaling of the peak height, which remains nonzero in the thermodynamic limit - further confirming the presence of long-range charge order.

Finally, we examine the Tomonaga-Luttinger liquid parameter \( K_\rho \), shown in Fig.~\ref{S14}(c) for \( U = 10t \), \( V = 10t \), and \( P = 2.5t \). The parameter \( K_\rho \) approaches zero with increasing system size, which is consistent with a charge-gapped, insulating state. The vanishing of \( K_\rho \) provides yet another indication that the CDW III phase is not metallic but exhibits strong charge localization.\\
\\
\\
\\
$*$manoranjan.kumar@bose.res.in\\
$\dagger$s.nishimoto@ifw-dresden.de\\
Last two authors contributed equally to this work.
\bibliography{sup_ref}